\documentclass[longauth]{aa}
\newcommand{\todo}{\ifmmode \text{\color{red}\Huge{\(\bullet\)}} \else {\color{red}{\Huge$\bullet$}}\fi}

\newcommand{  \hi       }{\ifmmode {\rm H}\,\textsc{i} \else H\,\textsc{i}\fi}
\newcommand \sn {\ensuremath{\mathrm{S/N}}}

\newcommand \gal {{M$\,$83}}
\newcommand \kms {{km$\,$s$^{-1}$}}

\newcommand \vlos {{$V_\mathrm{los}$}}
\newcommand \vsys {{$V_\mathrm{sys}$}}
\newcommand \vrot {{$V_\mathrm{rot}$}}
\newcommand \vrad {{$V_\mathrm{rad}$}}
\newcommand \vres {{$V_\mathrm{res}$}}
\newcommand \vlosobs {{$V_\mathrm{los, obs}$}}
\newcommand{\vlosmdl}{{$V_\mathrm{los, mdl}$}}

\newcommand \sigatom {\ensuremath{\mathrm{\Sigma_{HI}}}}
\newcommand \sigatomunit {\ensuremath{\mathrm{M_{\odot}\,pc^{-2}}}}

\newcommand \sig {\ensuremath{\mathrm{\sigma}}}
\newcommand \sigeff {\ensuremath{\mathrm{\sigma_{eff}}}}
\newcommand \sigmom {\ensuremath{\mathrm{\sigma_{\sqrt{\mathrm{mom2}}}}}}

\newcommand{\uv}{{$u{-}v$\,}}

\usepackage{graphicx}	
\usepackage{amsmath}	
\usepackage{multirow}
\usepackage{siunitx}
\usepackage{lipsum}
\usepackage{textcomp}
\usepackage{pifont}
\usepackage{xcolor}
\usepackage{txfonts}
\usepackage{csvsimple}
\usepackage{siunitx} 
\usepackage[utf8]{inputenc}
\usepackage{xparse}
\usepackage[pdfpagelabels=false]{hyperref}	
\hypersetup{colorlinks=true,linkcolor=blue,citecolor=blue,filecolor=blue,urlcolor=blue,}
\usepackage{gensymb} 

\newcommand{\ubonn}{Argelander-Institut f\"ur Astronomie, Universit\"at Bonn, Auf dem H\"ugel 71, 53121 Bonn, Germany}
\newcommand{\osu}{Department of Astronomy, The Ohio State University, 4055 McPherson Laboratory, 140 West 18th Ave, Columbus, OH 43210, USA}
\newcommand{\oan}{Observatorio Astronómico Nacional (IGN), C/ Alfonso XII, 3, E-28014 Madrid, Spain}

\newcommand{\rechen}{Astronomisches Rechen-Institut, Zentrum f{\"u}r Astronomie der Universit{\"a}t Heidelberg, M{\"o}nchhofstra{\ss}e 12-14, 69120 Heidelberg, Germany}
\newcommand{\ox}{Sub-department of Astrophysics, Department of Physics, University of Oxford, Keble Road, Oxford OX1 3RH, UK}

\newcommand{\anu}{Research School of Astronomy and Astrophysics, Australian National University, Canberra, ACT 2611, Australia}

\newcommand{\mpia}{Max Planck Institute for Astronomy, K\"onigstuhl 17, D-69117 Heidelberg, Germany}

\newcommand{\gent}{Sterrenkundig Observatorium, Universiteit Gent, Krijgslaan 281 S9, B-9000 Gent, Belgium}

\newcommand{\wyo}{Department of Physics \& Astronomy, University of Wyoming, Laramie, WY, 82071, USA}
\newcommand{\alb}{Dept. of Physics, University of Alberta, Edmonton, Alberta, Canada T6G 2E1}
\newcommand{\cape}{Department of Astronomy, University of Cape Town, Private Bag X3, Rondebosch 7701, South Africa}
\newcommand{\westv}{Department of Physics and Astronomy, West Virginia University, White Hall, Box 6315, Morgantown, WV 26506, USA}
\newcommand{\westvg}{Center for Gravitational Waves and Cosmology, West Virginia University, Chestnut Ridge Research Building, Morgantown, WV 26505, USA}
\newcommand{\eso}{European Southern Observatory, Karl-Schwarzschild Straße 2, D-85748 Garching bei M{\"u}nchen, Germany}
\newcommand{\ulyon}{Univ Lyon, Univ Lyon1, ENS de Lyon, CNRS, Centre de Recherche Astrophysique de Lyon UMR5574, F-69230 Saint-Genis-Laval France}
\newcommand{\cfa}{Center for Astrophysics, Harvard \& Smithsonian, 60 Garden St., 02138 Cambridge, MA, USA}
\newcommand{\hopk}{Department of Physics \& Astronomy, Bloomberg Center for Physics and Astronomy, Johns Hopkins University, 3400 N. Charles Street, Baltimore, MD 21218}
\newcommand{\theoheid}{Instit\"ut  f\"{u}r Theoretische Astrophysik, Zentrum f\"{u}r Astronomie der Universit\"{a}t Heidelberg, Albert-Ueberle-Strasse 2, 69120 Heidelberg, Germany}
\newcommand{\madrid}{Departamento de Fisica de la Tierra y Astrofisica \& IPARCOS, Facultad de CC Fisicas, Universidad Complutense de Madrid, 28040, Madrid, Spain}
\newcommand{\cool}{Cosmic Origins Of Life (COOL) Research DAO, coolresearch.io}
\newcommand{\eff}{Max-Planck-Institut für Radioastronomie, Radioobservatorium Effelsberg, Max-Planck-Strasse 28, Germany}
\newcommand{\nrao}{National Radio Astronomy Observatory, 1003 Lopezville Road, Socorro, NM 87801, USA}
\newcommand{\nraoc}{National Radio Astronomy Observatory, 520 Edgemont Rd, Charlottesville, VA 22903, USA}
\newcommand{\astron}{Netherlands Institute for Radio Astronomy (ASTRON),  Oude Hoogeveensedijk 4, 7991 PD Dwingeloo, Netherlands}
\newcommand{\kapeyn}{Kapteyn Astronomical Institute, University of Groningen, PO Box 800, 9700 AV Groningen, The Netherlands}
\newcommand{\uct}{Department of Astronomy, University of Cape Town, Private Bag X3, 7701 Rondebosch, South Africa}
\newcommand{\liverpool}{Astrophysics Research Institute, Liverpool John Moores University, 146 Brownlow Hill, Liverpool L3 5RF, UK}

\begin{document} 

   \title{Kinematic analysis of the super-extended HI disk of the nearby spiral galaxy M~83\thanks{Based on observations carried out with the Karl G. Jansky Very Large Array (VLA). The National Radio Astronomy Observatory is a facility of the National Science Foundation operated under cooperative agreement by Associated Universities, Inc.}}

   \author{Cosima Eibensteiner\inst{1}\fnmsep\thanks{\email{eibensteiner@astro.uni-bonn.de}}
          \and 
          Frank~Bigiel \inst{1}
          \and
          Adam~K.~Leroy \inst{2}
          \and
          Eric~W.~Koch
          \inst{3}
          \and
          Erik~Rosolowsky \inst{4}
          \and
          Eva Schinnerer \inst{5} 
          \and \\
          Amy Sardone \inst{2}
          \and 
          Sharon Meidt \inst{6}
          \and
          W. J. G de Blok \inst{7,8,9}
          \and
          David Thilker \inst{10}
          \and
          D. J. Pisano \inst{11,12,13}
          \and
          Jürgen Ott \inst{14}
          \and \\
          Ashley Barnes\inst{1,15}
          \and 
          Miguel Querejeta \inst{16}
          \and
          Eric Emsellem \inst{15,17}
          \and
          Johannes Puschnig\inst{1}
          \and
          Dyas Utomo \inst{18}
          \and
          Ivana~Be{\v{s}}li{\'c}\inst{1}
          \and \\
          Jakob den Brok\inst{1}
          \and 
          Shahram Faridani\inst{1}
          \and
          Simon~C.~O.~Glover \inst{19}
          \and
          Kathryn Grasha \inst{20}
          \and 
          Hamid Hassani \inst{4}
          \and 
          Jonathan~D.~Henshaw \inst{14,21}
          \and
          Maria J. Jim\'enez-Donaire \inst{15}
          \and
          Jürgen Kerp \inst{1}
          \and
          Daniel~A.~Dale \inst{22}
          \and 
          J.\ M.\ Diederik Kruijssen \inst{23}
          \and \\
          Sebastian Laudage \inst{1}
          \and 
          Patricia Sanchez-Blazquez \inst{24}
          \and
          Rowan Smith \inst{19}
          \and
          Sophia Stuber \inst{14}
          \and 
          Ismael Pessa \inst{14}
          \and \\
          Elizabeth J. Watkins \inst{25}
          \and 
          Thomas G. Williams \inst{26}
          \and
          Benjamin Winkel \inst{27}
          }
   \institute{\ubonn\
              \and
              \osu\
              \and
              \cfa\
              \and
              \alb\
              \and
              \mpia 
              \and
              \gent\
              \and
              \astron\ 
              \and
              \kapeyn\ 
              \and
              \uct\
              \and
              \hopk\
              \and
              \cape\
              \and
              \westv\
              \and
              \westvg\ 
              \and
              \nrao\ 
              \and
              \eso\
              \and
              \oan
              \and
              \ulyon
              \and
              \nraoc
              \and
              \theoheid\ 
              \and
              \anu\
              \and
              \liverpool\  
              \and
              \wyo
              \and 
              \cool\
              \and
              \madrid\
              \and
              \rechen\
              \and
              \ox\
              \and
              \eff\
              }

  \date{Received 25 October 2022 / Accepted 24 March 2023 }

  \abstract{We present new \hi\ observations of the nearby massive spiral galaxy \gal\, taken with the VLA at $21\arcsec$ angular resolution ($\approx500$~pc) of an extended ($\sim$1.5~deg$^2$) 10-point mosaic combined with GBT single dish data. We study the super-extended \hi\ disk of M83 (${\sim}$50~kpc in radius), in particular disc kinematics, rotation and the turbulent nature of the atomic interstellar medium. We define distinct regions in the outer disk ($r_{\rm gal}>$central optical disk), including ring, southern area, and southern and northern arm. We examine \hi\ gas surface density, velocity dispersion and non-circular motions in the outskirts, which we compare to the inner optical disk. We find an increase of velocity dispersion ($\sigma_v$) towards the pronounced \hi\ ring, indicative of more turbulent \hi\ gas. Additionally, we report over a large galactocentric radius range (until $r_{\rm gal}{\sim}$50~kpc) that $\sigma_v$ is slightly larger than thermal (i.e. $>8$\kms\ ). We find that a higher star formation rate (as traced by FUV emission) is not always necessarily associated with a higher \hi\ velocity dispersion, suggesting that radial transport could be a dominant driver for the enhanced velocity dispersion. We further find a possible branch that connects the extended \hi\ disk to the dwarf irregular galaxy UGCA\,365, that deviates from the general direction of the northern arm.
  Lastly, we compare mass flow rate profiles (based on 2D and 3D tilted ring models) and find evidence for outflowing gas at r$_{\rm gal}$ $\sim$2~kpc, inflowing gas at r$_{\rm gal}$ $\sim$5.5~kpc and outflowing gas at r$_{\rm gal}$ $\sim$14~kpc. We caution that mass flow rates are highly sensitive to the assumed kinematic disk parameters, in particular, to the inclination.}
  

   \maketitle
%
\renewcommand{\equationautorefname}{Eq.}
\renewcommand{\sectionautorefname}{Section}
\renewcommand{\subsectionautorefname}{Section}
\renewcommand{\subsubsectionautorefname}{Section}

\section{Introduction}

Massive, extended atomic gas reservoirs that often surround spiral galaxies usually extend far beyond the inner optical disk ($2{-}4 \times r_{25}$, where $r_{25}$ is the optical radius; e.g.\ \citealt{Wang2016}). These reservoirs may eventually serve as the fuel for star formation in the inner disk and may represent a key component for facilitating accretion from the circumgalactic medium (CGM) and cosmic web (see e.g.\ review by \citealt{Sancisi2008}). However, the details of star formation in outer disks, the origin, and nature of turbulence (traced by, for example, \hi\ velocity dispersion), and perhaps most importantly radial gas flows all remain topics in need of more study.

Observing the extent of neutral atomic hydrogen (\hi) gas is important as it traces the kinematics in disk galaxies which gives further insights into the understanding of galaxy evolution. The 21-cm line emission of \hi\ is often used to examine either the disc kinematics and rotation on large scales \citep[e.g.][]{deBlok2008,Heald2011,Schmidt2016,Oman2019,DiTeodoro2021}, or the turbulent nature of the interstellar medium (ISM) on small scales (e.g.\ \citealt{Tamburro2009,Ianjamasimanana2012,Ianjamasimanana2015,Mogotsi2016,Koch2018}). 
In the first case, \hi\ traces the process of gas accreting from the intergalactic medium (IGM) that flows through diffuse filamentary structures into the CGM of galaxies and further onto the galaxy disk (e.g.\ \citealt{Keres2005}). The gas entering the halo remains cool as it falls onto the disk (i.e.\ the cold mode scenario in \citealt{Keres2005}). This cold accretion dominates in the lower \hi\ density environments \citep[e.g.][]{White1991,Keres2005,Tumlinson2017}. In the second case, \hi\ is used to analyze the velocity profile and line widths (i.e.\ \hi\ velocity dispersion) to interpret the thermal states of the optically thin warm neutral medium (WNM, as one of the two phases predicted by models from, e.g. \citealt{Field1969,Wolfire1995,Wolfire2003,Bialy2019}). This is found to be valid down to 100~pc scales (e.g.\ \citealt{Koch2021}), for example, from emission and absorption studies in the LMC and SMC (\citealt{Stanimirovic1999,Jameson2019}).   

The exact amount of \hi\ gas in the outskirts of nearby galaxies was, for example, analyzed in \cite{Pingel2018} and \cite{Sardone2021}. \cite{Pingel2018} analyzed and compared four galaxies out of 24 total sources of the Hydrogen Accretion in LOcal GAlaxies Survey -- HaloGAS (see \citealt{Heald2011} for the survey paper) with Green Bank Telescope (GBT) observations. They found that the \hi\ mass fraction below \hi\ column densities of $N_{\rm HI} = 10^{19}\,{\rm cm^{-2}}$ (i.e.\ diffuse \hi\ gas) is on average 2$\%$. Furthermore, their GBT observations of NGC\,925 reveal a detection of ${\sim}20\%$ more \hi\ than observations done with the VLA as part of The HI Nearby Galaxy Survey (THINGS; \citealt{Walter2008}). This underscores, among other aspects, that the THINGS VLA interferometric observations require a correction for missing short-spacing observations. \cite{Sardone2021} studied 18 out of a total of 30 nearby disk and dwarf galaxies of the MeerKAT \hi\ Observations of Nearby Galactic Objects; Observing Southern Emitters -- MHONGOOSE (see \citealt{deBlok2016} for the survey paper), and found that 16 out of 18 galaxies have 0.02-3 times additional \hi\ mass outside of their optically bright disks.

\begin{figure*}
    \centering
    \includegraphics[width=1.0\textwidth]{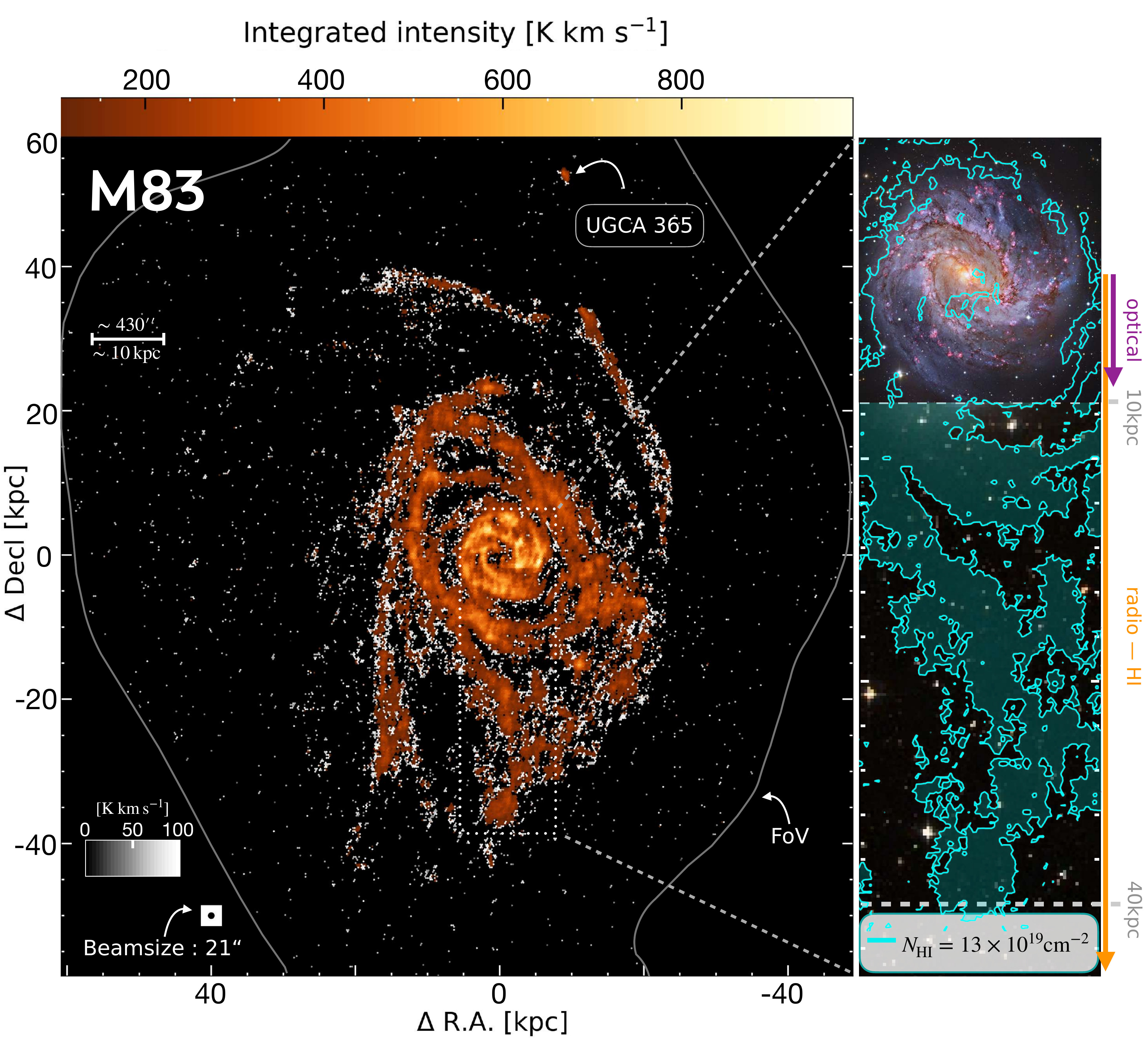}
    \caption{{The integrated intensity map (moment 0) for \hi\ emission across the the disk of \gal\ at a resolution of 21$\arcsec$.} The black circle in the lower left corner marks the beam size of 21$\arcsec$ (${\approx}$500\,pc). The black to white colorbar indicates integrated intensities from $0-100$~K~km~s$^{1}$ and the orange to yellow integrated intensities above $101$~K~km~s$^{1}$. We denote the companion galaxy UGCA$\,$365 which in projection is ${\sim}$55$\,$kpc away from the center of \gal. To the right, we show the enclosed optical disk (r$_{25}{\sim}8$~kpc) overlaid with \hi\ column density contour ($N_{\rm HI} = 13\times10^{19}$\,cm$^{-2}$) extending over a radius of ${\sim}40$~kpc. For visualization reasons, we show unfilled $N_{\rm HI}$ contours for the high resolution optical image. Beyond 10~kpc we show filled $N_{\rm HI}$ contours. The white line surrounding \gal\ shows the field of view (FoV) (i.e. the full mosaic coverage) of the VLA observation. (optical image credits: CTIO/NOIRLab/DOE/NSF/AURA, M. Soraisam; Image processing: Travis Rector, Mahdi Zamani $\&$ Davide de Martin; low resolution background > 10~kpc: DSS2)}
    \label{fig:mom0}
\end{figure*}
\begin{figure*}
    \centering
    \includegraphics[width=0.85\textwidth]{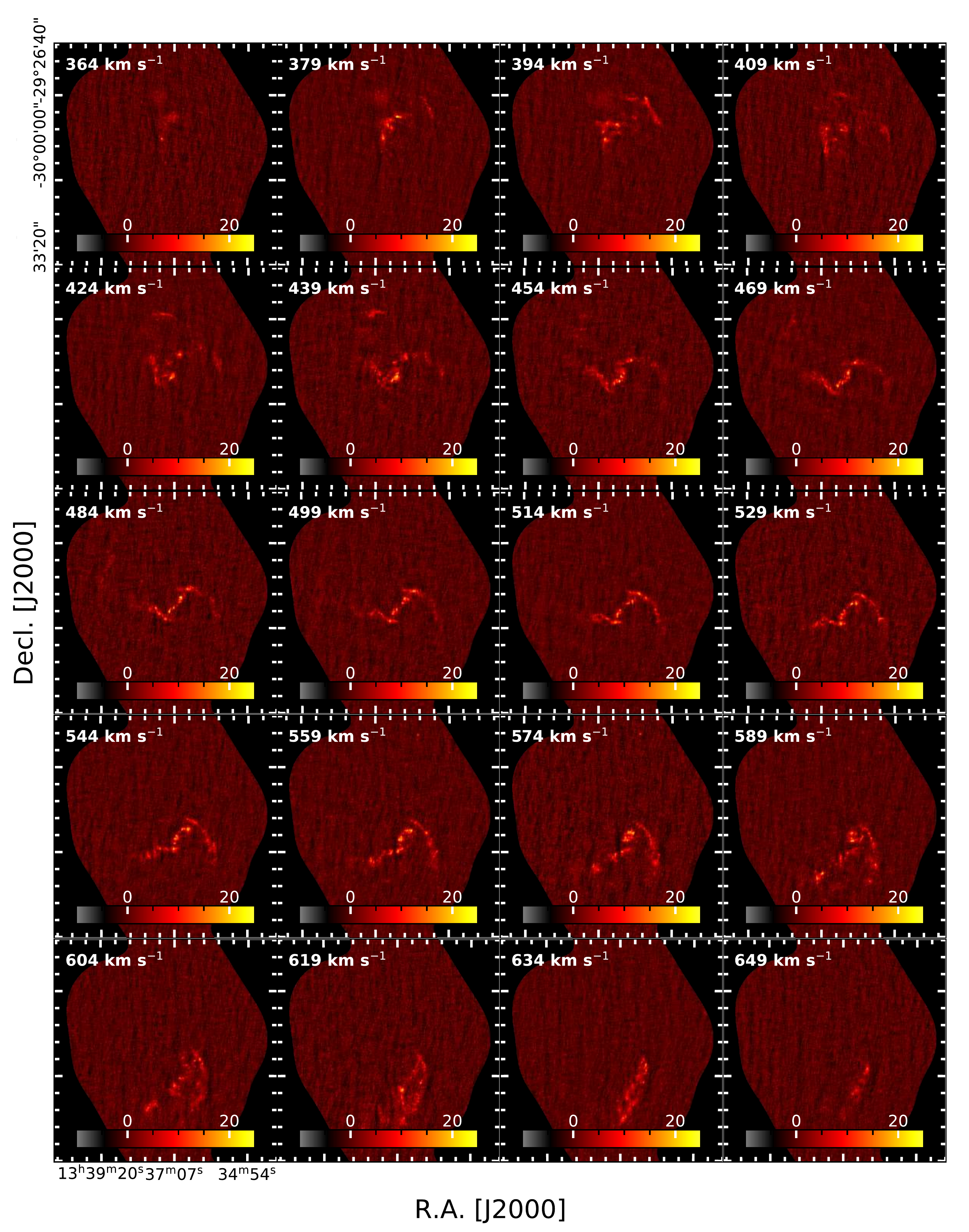}
    \caption{Channel maps of of our VLA+GBT data. The line-of-sight velocity of the shown channels is displayed on the upper left corner of each panel and the colorbar is shown in units of K. For visual purpose we only show every third $5.0$ \kms\ channel (i.e.\ we do not integrate over channels). The companion galaxy UGCA\,365 is detected in the velocity channels $559$ \kms\ and $574$ \kms. We show all channels in the Appendix (see \autoref{appendix:add_figures}).}
    \label{fig:chanmaps}
\end{figure*}
The nearby ($D=5.16$\,Mpc) grand-design spiral galaxy \gal\ (also known as NGC\,5236, see \autoref{Tab: Properties}) is massive, favorably oriented ($i=48^\circ$), and associated with active star formation (see \autoref{Tab: Properties}), which has made it a classic target for several studies of the \hi\ in outer portions of disk galaxies \citep[e.g.][]{Huchtmeier1981,Tilanus1993,Miller2009}. Moreover, \gal\ was one of the first galaxies where the tilted ring analysis was performed (\citealt{Rogstad1974}, see their fig.\ 8). Sensitive, wide-area \hi\ imaging of \gal\ reveal its highly extended, structured disk, including a northern spiral arm-like structure \citep[e.g][]{Heald2016,Koribalski2018}. However, previous high-resolution interferometric observations \citep[see e.g][]{Walter2008,Bigiel2010} have been restricted to a single central pointing with the Very Large Array ($0.5^\circ$ primary beam). In this paper, we present a 10-field mosaic using the VLA 
with an angular resolution of 21$\arcsec$($\sim 500\,$pc) corrected for short-spacing using GBT observations (see \autoref{sec:imaging}).

The aim of this paper is to study the large and small-scale kinematics of the \hi\ in \gal's outskirts ($r_{\rm gal}>r_{\rm 25}$), where we look at individually defined regions and compare their \hi\ properties to the central (optical) disk. We search for deviations from pure circular motions and asymmetries in the outskirts. We examine radial trends in velocity dispersion and environmental differences between the defined regions.   
We will further highlight the effects of different kinematic parameters (based on 2D and 3D tilted ring models) on average radial mass flow rates 
and tilted ring modeling in general.

The paper is structured as follows: in \autoref{sec:Observation} we describe how the VLA observations were taken, calibrated, and imaged, along with how we convert observational measurements to physical quantities. \autoref{sec:Results} presents the distribution of \hi\ across \gal\ , its velocity field, and environmental differences of \hi\ (\hi\ gas surface density, \sigatom) and kinematic parameters (\hi\ velocity dispersion, line of sight velocities and residual velocities). In \autoref{sec:Discussion} we discuss the environmental dependence of the observed \hi\ velocity dispersion along with the northern extend arm of \gal\ and \gal \textquotesingle s possible interaction with a nearby dwarf irregular galaxy UGCA$\,$365. In \autoref{sec:massflow} we show upper limits for average radial mass flow rates (based on 2D and 3D tilted ring models) and discuss the limitations and compare to mass flow rates from the literature. 

\begin{table}
\begin{center}
\caption{Properties of \gal, NGC$\,$5236}
\label{Tab: Properties}
\begin{tabular}{lcc}
\hline \hline
Parameter & Value        &   Notes \\ \hline
Morphology           & SAB(s)c       &     (1)       \\
R.A. (J2000)         &   204.2538 deg &  (2)  \\
Decl. (J2000)        &  -29.8658 deg &   (2)      \\
Distance        &  5.16$\pm$0.41    &  (3) \\
Linear scale & $25$~pc$/$arcsec & \\
D$_{25}$        & 11.7 arcmin    & (4) \\
r$_{25}$        & ${\sim}8$~kpc     & (4)       \\
log$_{10}$ SFR &   $0.62\,M_{\odot}\,{\rm yr^{-1}}$        &    (5)  \\
log$_{10} \,L_{\rm CO}$ &   $8.84\,{\rm K\,km\,s^{-1}\,pc^{2}}$        &    (6) \\
Adopted parameters: & & \\
    \enspace Inclination      & 48$\degr$      &  (7) \\
    \enspace P.A. & 225$\degr$& (7) \\
    \enspace \vsys\     & 510 \kms\ &  (7) \\
    \enspace \vrot\ & 62 \kms\ & (7) \\
\hi\ mass of: &         & \textit{This work}        \\
 \enspace central disk  &    1.1 $\times$ 10$^{9}$ M$_{\odot}$     & (8)       \\
 \enspace ring  &           1.9 $\times$ 10$^{9}$ M$_{\odot}$      &       \\
 \enspace southern area  &     1.8 $\times$ 10$^{9}$ M$_{\odot}$ &       \\
 \enspace southern arm  &     0.7 $\times$ 10$^{9}$ M$_{\odot}$    &       \\
 \enspace northern arm  &     0.5 $\times$ 10$^{9}$ M$_{\odot}$    &       \\
\hline 
\end{tabular}
\end{center}
 \begin{minipage}{0.95\columnwidth}
        \vspace{1mm}
        {\bf Notes:} (1):  \cite{deVaucouleurs1991}. (2): \cite{Wang2016}. (3): 
        \cite{Karachentsev2007}. (4): \cite{Walter2008}. We convert to physical units using the distance mentioned in this table. (5): Adopted from \cite{Leroy2019} using the FUV and WISE4 Band. (6): CO luminosity using CO(2-1) from the PHANGS-ALMA survey, \cite{Leroy2021}  (7): Kinematic parameters from \cite{Heald2016} that we use in this work and describe in \autoref{sec:adopted_tilted_ring_model}. We quote here the ones for the central tilted ring. (8): The total \hi\ mass of each region is calculated including single dish observations. 
    \end{minipage}
\end{table}

%
%
\section{Observations, Data Reduction and Products} 
\label{sec:Observation}
\subsection{Observations}
The observations for \gal\ were carried out at the VLA. We obtained a 10-point mosaic 
using the dual-polarization L-band of the VLA to map the 21-cm emission of neutral hydrogen over the entire super-extended disk of the nearby galaxy \gal\ (see \autoref{fig:mom0} and \autoref{fig:chanmaps}). Most of the data were taken from February 2014 through January 2015 for $\sim$40h over the course of 22 runs (project codes: 13B-196, 14B-192, PI: F.~Bigiel). We used the VLA in the three hybrid configurations D north C (DnC), C north B (CnB), and B north A (BnA), to ensure good $u$-$v$ coverage despite M83's southern declination. Based on the THINGS observing strategy, we chose a similar time split for these configurations (see \autoref{tab:observation}). We chose a velocity resolution of 0.5 km~s$^{-1}$ and a bandwidth of 4~MHz corresponding to 860~km~s$^{-1}$, which is well suited to resolve the \hi\ line and cover the full range of 21-cm emission from the galaxy with enough bandwidth for continuum subtraction. 
At the beginning of each observing session, we observed for $\sim$10 min the bandpass calibrator 3C286. Observations were then set up so that two of the ten mosaic pointings were observed in succession for 10 min each, followed by the gain calibrator (J1331-2215). During each session, the gain calibrator was observed 6 times. We show a summary of our observations in \autoref{tab:observation}.

\subsection{Reduction}

The VLA pipeline implemented in \texttt{CASA} (\citealt{McMullin2007}; version: 5.4.2-8.el6) was used without Hanning smoothing, as recommended by the VLA pipeline guide\footnote{\url{https://science.nrao.edu/facilities/vla/data-processing/pipeline}} for each of the science blocks. The flagging summary showed us that one to a maximum of two antennas were completely flagged for some science blocks. As a next step, we reduced the data to include only the spectral window (spw) where the HI emission is located (rest-frequency $\sim$ 1.420 GHz) using the CASA task \texttt{split}. This was followed by producing various diagnostic plots with a focus on the calibrator sources and applying additional data flagging (e.g.\ spikes of radio frequency interference or baselines) by hand using the CASA task \texttt{flagdata}. 

Then we ran a standard calibration script where we calibrate the flux, bandpass, and gain using \texttt{gaincal}. As the calibrators did not show any discrepancies, we merged all science blocks via CASA function \texttt{concat}. This results in a measurement set containing only the \hi\ spw with the target field scans (10 pointing). During our first run of a test \texttt{tclean}, we noticed horizontal and vertical stripes in the channels around the local standard of rest velocity of 380 \kms. We carried out several additional rounds of manual flagging to identify and remove corrupted baselines. We inspected amplitudes and phases for the data from all baselines by eye, finding some corrupt baselines that we flagged. 

\subsection{Imaging}
\label{sec:imaging}
To create 'images' of our VLA observation we used the CASA task \texttt{tclean} that combines multiple arrays and deconvolved using the \texttt{multiscale} clean algorithm \citep{Cornwell2008}. We use the \texttt{mosaic} gridding algorithm and weigh the \uv\ data according to the Briggs scheme with robustness parameter $r=0.5$ which balances spatial resolution and surface brightness sensitivity. We tried several settings (for example, (i) restoringbeam=15$\arcsec$ with cell=5$\arcsec$ (ii) restoringbeam=21$\arcsec$ with cell=7$\arcsec$) and find that 
a resolution of 21$\arcsec$ (restoringbeam=21.0) and a spectral resolution of 5~\kms\ (see \autoref{tab:observation_properties}) provides a good compromise between resolution and noise. Since our observations have more integration time and substantially better surface brightness sensitivity in the D-configuration data, our choice of weighting naturally produces a synthesized beam size that is comparable to the D-configuration beam (46$\arcsec$, see \autoref{tab:observation_properties}).
We restricted the deconvolution using a clean mask derived using CASA's automasking scheme (\citealt{Kepley2020}) with the following sub-parameters: sidelobethreshold=0.75, noisethreshold=3.0, lownoisthreshold=2.0, negativethreshold=0.0, minbeamfrac=0.1, growiterations=75. In order to increase the signal-to-noise (\sn) to be sensitive to emission even where the line is faint, we average several channels, resulting in a spectral resolution of $\Delta_{v}=5.0 \,$\kms. The typical rms per channel observed is $\approx3$~mJy~ beam$^{-1}$. 

The \hi\ in \gal\ is extended compared to the primary beam of the VLA. As a result, subsequent to deconvolution, the interferometric VLA data need to be combined with single-dish data to correct for the insensitivity of the interferometer to extended emission. We used Green Bank Telescope single dish data that was obtained as part of the GBT-THINGS project (project GBT11A-055) and similar to the maps in \citet{Pisano2014}. The GBT data were taken on 6, 12, and 26 March 2011. Either 3C147 or 3C295 was observed as a primary flux calibrator before mapping 4 square degrees around \gal. We derived a T$_{\rm cal}$ value of 1.55~K in both polarizations that we used for all observations.  We used the GBT spectrometer with a 50~MHz bandwidth for observing in a frequency-switching mode with a 10~MHz throw while making a basket-weave map in RA and Dec. While data were taken with frequency-switching, we used the edge 4 integrations as an "off" position to calibrate the maps. These correspond to 6.7 arcmin on each edge. Since the extent of \gal\ is less than 2 deg, and the map shows no subtracted \hi\ emission, we find that these are clean "off" positions. A second order polynomial was fit to the spectra over an emission free region to remove residual baseline structure. Data were boxcar smoothed to 5.15 \kms\ resolution before being imaged with the \texttt{SDGRD} task in AIPS. A fourth order polynomial was removed from the final cube. 

\begin{figure}
    \centering
    \includegraphics[width=0.45\textwidth]{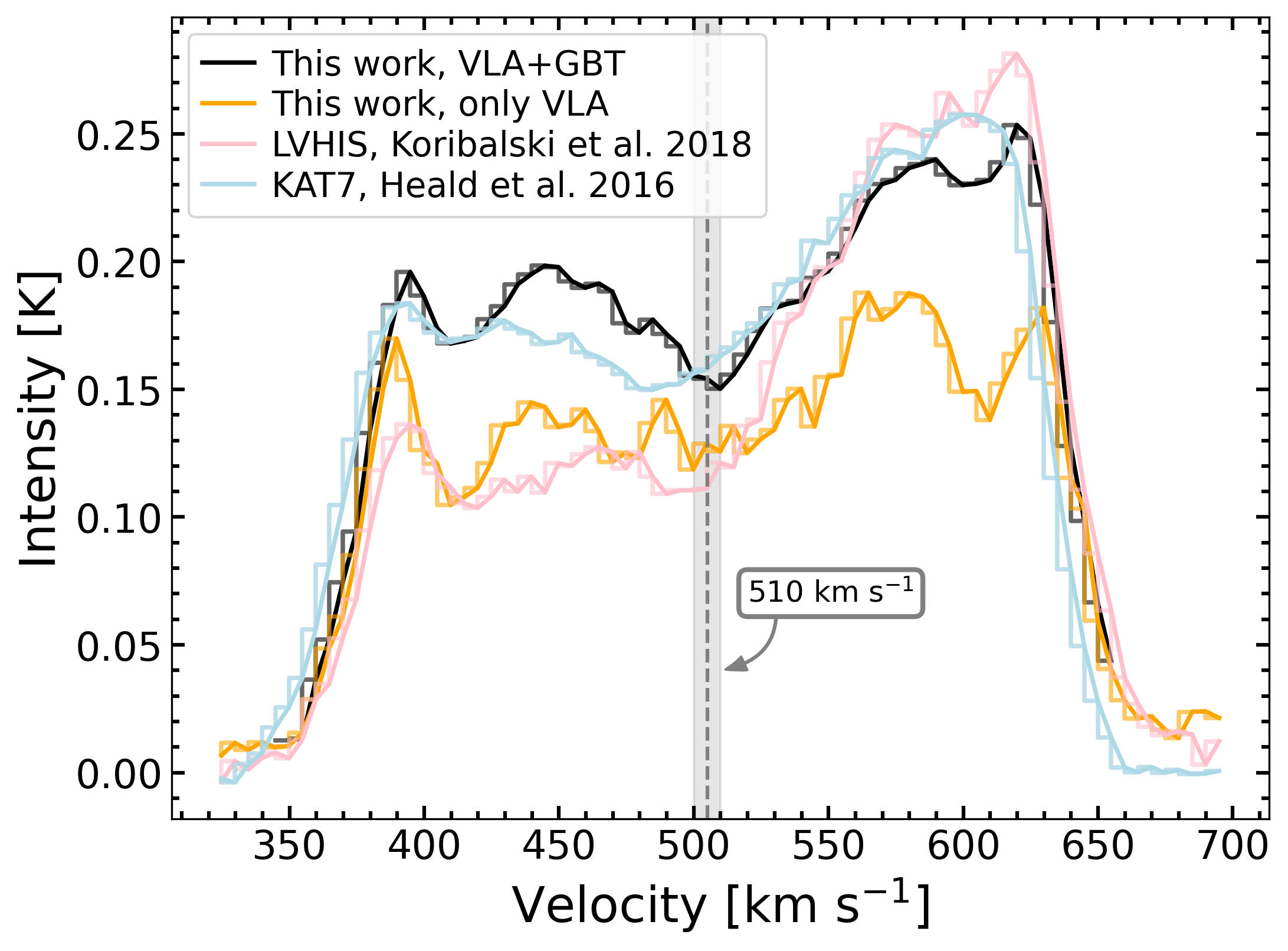}
    \caption{The global \hi\ profile of \gal\ from our VLA observations (orange) and the results of feathering the VLA with the GBT observations (black). The systemic velocity appropriate for the centre of the galaxy is indicated with the vertical grey dashed line at $510$~\kms\ (see also \autoref{Tab: Properties}). We also show \hi\ observations of the same FoV of \gal\ from LVHIS (\citealt{Koribalski2018}) and KAT7 (\citealt{Heald2016}) that we have sampled on the same spectral axis grid at their native angular resolution.}
    \label{fig:spec}
\end{figure}

We combined the GBT map with the mosaic interferometer data using the \texttt{CASA} task \texttt{feather}. Some of the emission is very extended and must be recovered by feathering, so the pixel statistics often reflect this extended emission. Before performing this task, we used the \texttt{uvcombine} python package (\citealt{Koch2022}) to find the correct single dish scaling factor (sdfactor). We used a sdfactor of $1.0$ (see \autoref{fig:feather_compare}). We regard lines of sight separated by more than a VLA synthesized beam as being statistically independent.  However, because of the feathering process, the noise in the GBT data may lead to weak but spurious correlations up to scales approaching the GBT beam of 523$\arcsec$.  Since the surface brightness of the noise in the GBT map is small compared to the VLA data and our results do not depend heavily on the statistical independence of the VLA data, we note this as a caution and proceed. We show properties of the feathered data in \autoref{tab:observation_properties}. In \autoref{fig:spec} we show the spectra of the VLA and the feathered VLA (VLA+GBT) cubes after we converted them to units of Kelvin. We find a total flux for the VLA cube of $5.2\times 10^{4}$~K and $6.6\times 10^{4}$~K for the feathered one (VLA+GBT). This emphasizes that single dish data are needed even for interferometric observations including compact configurations. For a comparison we also show in \autoref{fig:spec} the spectra over the same field of view from The Local Volume H I Survey(LVHIS; \citealt{Koribalski2018} f with ATCA at ${\sim}$113$\arcsec{\approx}$2.62~kpc scales) and KAT7 observations (seven-dish MeerKAT precursor array, \citealt{Heald2016} at ${\sim}$230$\arcsec{\approx}$5.33~kpc scales), which were sampled on the same spectral axis grid. We see that the KAT7 agrees well with the VLA+GBT spectra while LHVIS only agrees on the receding side (i.e.\ $>510$\kms). The spectra of VLA+GBT highlights a significant fraction of emission that we see with the addition of the GBT observation.

\begin{table*}
\centering
\caption{Summary of our VLA $\hi$ observation across \gal. }
\label{tab:observation}
\begin{tabular}{lccc|cc|ccc|ccc}
\hline \hline
\multicolumn{1}{c}{}  &
  $\nu_{\rm rest}$ &
  \begin{tabular}[c]{@{}c@{}}project\\ codes\end{tabular} &
  \begin{tabular}[c]{@{}c@{}} sessions\end{tabular} &
  \begin{tabular}[c]{@{}c@{}}gain/phase\\ calibrator\end{tabular} &
  \begin{tabular}[c]{@{}c@{}}bandpass/flux\\ calibrator\end{tabular} &
  config. & {$\theta_{\rm HPBW}$} &
  \begin{tabular}[c]{@{}c@{}}obs. time \\ per config.\end{tabular} &
  \begin{tabular}[c]{@{}c@{}} total \\ obs. \\ time\end{tabular} & 
  \begin{tabular}[c]{@{}c@{}}total \\ on\\ source\end{tabular}\\
\multicolumn{1}{c}{\multirow{-2}{*}{}} & [MHz] & & & & & & [$\arcsec$] & [min] & [min] & [min]   \\
& (1) & (2) & (3) & (4) & (5) & (6) & (7) & (8) & (9) & (10)  \\ \hline
\hi\   & 1420.4058 & 13B-194 & 17 &J1311-2216     &  1331+305=3C286         & BnA   & 4.3 & 378  &  2406  & 1444  \\
&  & and & & and &  & CnB  & 14 & 516  & &       \\
&  & 14B-192 & 5 & J1316-3338&  & DnC & 46 & 1512 & &      \\
\hline 
\end{tabular}
 \begin{minipage}{2.0\columnwidth}
        \vspace{1mm}
        {\bf Notes.} Column (1): Observational frequency. (2--3): The project codes and the corresponding number of observation sessions (PI: F. Bigiel). (4): The sources we used to calibrate the gain (amplitude and phase). (5): The source we used to perform the bandpass (flux and delay) calibration. (6--10): We quote here the outputs using \texttt{analysisUtils} after we have flagged our data:  (7): estimates of the synthesized beamwidth taken from the manual for the VLA observational status summary 2013B \footnote{\url{https://science.nrao.edu/facilities/vla/docs/manuals/oss2013B/performance/resolution}} (8): Time of the observation per configuration. (9): Total observation time. (10): Total on source time. 
    \end{minipage}
\end{table*}

\begin{table*}
\centering
\caption{Properties of our imaged and feathered dataset.}
\label{tab:observation_properties}
\begin{tabular}{lccc|ccccccc|c}
\hline \hline
\multirow{2}{*}{}  & \multicolumn{2}{c}{Beam} & $\Delta\nu_{\rm chan}$ &$I_{\rm HI}$ &$\sigma{\rm_{HI}}$ & S/N & $T_{\rm peak}$ & Noise & \vlos & $\sigma{_{V{\rm los}}}$ & \multicolumn{1}{c}{\begin{tabular}[c]{@{}c@{}}Ratio with\end{tabular}} \\
& [$\arcsec$] & [pc] & [km~s$^{-1}$] &  [K~km~s$^{-1}$] & [K~km~s$^{-1}$] & & [K] & [K~km~s$^{-1}$]     & [km~s$^{-1}$] & [km~s$^{-1}$] & FUV \\ 
& (1) & (2) & (3) & (4) & (5) & (6) & (7) & (8) & (9) & (10) & (11)\\ \hline
\multicolumn{1}{l}{\hi} & 21 & 483 & 5 & 991 & 32 & 31 & 30 & 1.6 & 570 & 1.3 & 9.5$\times 10^{-4}$\\
\hline 
\end{tabular}
 \begin{minipage}{2.0\columnwidth}
        \vspace{1mm}
        {\bf Notes.} Column (1--2): The size of the beam in angular and linear scales (adopting a distance of 4.75~Mpc). (3): The channel width, i.e. spectral resolution. (4--11): We quote the quantities for the sight line with the highest \hi\ integrated intensities, i.e. for an aperture of $21\arcsec \approx 500$~pc  (see \autoref{tab:regions} for the mean, the 16th and 84th percentiles of some of these quantities): (4): Integrated intensity. (5): Uncertainty of integrated intensity (6): Signal to noise. (7): Peak temperature of the spectrum. (8): Root mean square (rms) noise. (9): Line of sight velocity, i.e. first-moment. (10): Uncertainty of the first-moment.  (11): Ratio of $I_{\rm HI}$ with FUV which results in units of K km s$^{-1}$/(mJy arcsec$^{-2}$). These are GALEX FUV observations that have been published in \cite{Bigiel2010}. 
    \end{minipage}
\end{table*}

\subsection{Data products}
\label{sec:Data_products}

After feathering, the data are resampled onto a hexagonal grid with a grid size matching the beam size. This ensures that the resampled measurements are \textit{nearly} mutually independent. This results in 50,574 sightlines separated by a beam size. To improve the signal-to-noise ratio (\sn) we applied a masking routine based on the methodology introduced by \cite{Rosolowsky2006}. First we identified pixels with a high \sn\ (\sn\ $\geq$ 4) and low \sn\ (\sn\ $\geq$ 2). As a next step the identified high S/N regions are iteratively grown to include adjacent moderate S/N regions as defined by the low S/N mask. In this way, we recover the more extended 2-sigma detection that belongs to a 4-sigma core. Only pixels detected with a S/N $\geq$ 3 (where the signal represents the integrated intensity; see next paragraph) are used in subsequent analysis. In this work we will make use of the following data products:

\begin{enumerate}
    \item The \textit{integrated intensity map} is created from the masked data cube by integrating along the velocity axis for each of the individual sightlines $v$ multiplied by the channel width $\Delta v_{\rm chan}$ of  5 \kms:
    \begin{equation}
    I_{\rm HI} \, [{\rm K \, km \, s^{-1}}] = \sum_{n\rm_{chan}} I_v \, [{\rm K}] \, \Delta v_{\rm chan} \, [\rm km \, s^{-1}]
    \end{equation}
    The uncertainty is calculated taking the square-root of the number of the included channels along a line-of-sight multiplied by the 1$\sigma$ root-mean-squared value of the noise and the channel width: $\sigma_\mathrm{HI} = \sqrt{n\rm_{chan}} \times \sigma_{\rm rms} \times \Delta v_{\rm chan}$. We calculate $\sigma_{\rm rms}$ over the signal-free part of the spectrum using the \texttt{astropy} (\citealt{Astropy2013,Astropy2018}) function \texttt{mad\_std} that calculates the median absolute deviation and scales it by a factor of 1.4826. This factor results from the assumption that the noise follows a Gaussian distribution. \\

    For further analysis\footnote{We use this \sn\ $>3$ criterion to construct \vlos, \vres, \sig\ and \sigatom.} we only focus on significant detections of $\sn\ = I_{\rm HI}/\sigma_{\rm HI} >3$ , resulting in 5,539 sightlines separated by a beam size.\\ 
    
    \item We convert the \hi\ 21-cm intensity ($I\mathrm{_{HI}}$) to the \hi\ \textit{gas surface density}, \sigatom, via: 
    \begin{equation}
        \Sigma_\mathrm{HI}~\mathrm{[M_{\odot}~pc^{-2}]} = 0.015~I\mathrm{_{HI}}~ \mathrm{[K~km~s^{-1}]} ~ \cos{(i)}.
    \end{equation}
    This conversion was used by \citet[among many others]{Bigiel2010_Ineff} and results in a hydrogen mass surface density and neglects heavy elements. The $\cos(i)$ factor corrects for inclination (see \autoref{Tab: Properties}). 
        We derive the \hi\ \textit{column density} via: 
    \begin{equation}
   N_{\mathrm{ \hi}}~[\mathrm{cm^{-2}}] = 1.82\times10^{18}~ {I\mathrm{_{HI}}~\mathrm{[ K~km~s^{-1}]}}.
    \end{equation}
    
    \item  Velocity fields aim to give an accurate characterization of the dynamics in a galaxy. For this purpose, each pixel is assigned a velocity that represents the average line-of-sight velocity of the gas. However, this is not trivial and multiple approaches are used in the literature that differ from each other by some advantages and disadvantages (see \citealt{deBlok2008} for a discussion on \textit{First-moment}, \textit{Peak velocity fields}, \textit{Gaussian profiles}, \textit{Multiple Gaussian profiles}, and \textit{Hermite h3 polynomials} methods). In this work we take for the \textit{observed velocity field} the first-moment as follows:
    \begin{equation}\label{eq:1mom}
        V_{\rm los} [{\rm km \, s^{-1}}] =  \frac{\sum{{I_{\nu} [{\rm K}] \, v~[{\rm km \, s^{-1}}}]~\Delta v_{\rm chan} [{\rm km \, s^{-1}}]}}{\textit{I}_{\rm HI} [\rm K~km~s^{-1}]}.
    \end{equation}
    
    \item We use the width of the \hi\ emission line to trace the \textit{velocity dispersion} ($\sigma_{v}$) along each line-of-sight. Several methods exist to estimate the width of an emission line: fitting the line profiles with a simple Gaussian function or Hermite polynomials or calculating the second moment. 
    
    In this work, we calculate the velocity dispersion $\sigma_v$ following the (i) "effective width", \sigeff\, approach used in, for example, \cite{Heyer2001}, \cite{Leroy2016} and, \cite{Sun2018}: 
    
    \begin{equation}\label{eq:sigma}
       \sigma_{\mathrm{eff}}~\mathrm{[km~s^{-1}]}=  \frac{\textit{I}_{\mathrm{HI}}~\mathrm{[ K~km~s^{-1}]}}{T_{\rm peak}~\mathrm{[K]}~\sqrt{2 \pi}}.
    \end{equation}
    The velocity-integrated intensity is divided by the peak brightness temperature, $T_\mathrm{peak}$. This is converted to the standard, intensity-weighted, velocity dispersion, by dividing with $\sqrt{2 \pi}$ -- the conversion constant for a Gaussian profile. This definition of $\sigma$ has the advantage that it is less sensitive to noise but will mis-characterize line profiles that significantly deviate from a single Gaussian.
    
    We do not subtract the finite channel width (i.e. line broadening caused by the instrument) as it is done, for example, for extragalactic CO observations (e.g. \citealt{Sun2018}). The usual finite correction factor will overcorrect given that 5 \kms\ $\approx$ the WNM thermal width.

    We also calculate $\sigma_v$ with (ii) the square root of the second moment: 
    \begin{equation}\label{eq:sigma_mom2}
    \begin{split}
         \sigma_{\sqrt{\mathrm{mom2}}}~\mathrm{[km~s^{-1}]} =  \left\{ \frac{\sum I_{\nu}\mathrm{~[K]~(v~[km \, s^{-1}]}-V_{\rm los}~[{\rm km \, s^{-1}}])^{2}}{\sum I_{\nu}~\mathrm{[K]}}\right\}^{1/2} .
    \end{split}
    \end{equation}
    We compare (i) \sigeff\ and (ii) \sigmom\ in \autoref{sec:velocitydispersion}.
    
\end{enumerate}

\subsection{Radial profiles -- binning and stacking}
\label{sec:binning_stacking}
Throughout this work, we make use of radial profiles, binning in galactocentric rings of 1~kpc width (roughly twice the beam size of 500~pc). Each point in those profiles thus represent the average within a given ring defined by the structure parameters (\autoref{Tab: Properties}).

In order to average (``stack'') spectra within a given ring, we align the spectra to the peak velocity (see e.g.\ \citealt{Jimenez-Donaire2019}, \citealt{Beslic2021}, or for \hi\ related science \citealt{Koch2018} who discussed different stacking approaches i.e., different definitions of the line center $V_{\rm rot}$, $V_{\rm cent}$, and $V_{\rm peak}$). We create stacks again in ${\sim}1$~kpc wide galactocentric rings. 

\subsection{Tilted ring kinematics}
\label{sec:methods}

\hi\ rotation curves are most commonly derived from velocity fields. The quantity that is accessible through observations is the line-of-sight velocity (\vlos). To infer the rotational velocities of the gas in the disk from measurements of \vlos, we have to utilize a specific model that can be fitted to the data. 
    
The `tilted-ring' approach describes a galaxy by a set of rings, each with their own inclination $i$, position angle, systemic velocity \vsys, center position ($x_0$, $y_0$), and rotation velocity \vrot. Under the assumption that the gas moves in circular orbits within each ring, we can then describe \vlos$_{\mathrm{, obs}}$ for any position (x, y) on a ring with radius $r$ as:
\begin{equation}
    \begin{split}
        V_{\rm los, obs}(x, y) = V_{\rm sys} + V_{\rm rot}(r) \sin(i) \cos(\theta) \\
        + V_{\rm rad}(r) \sin(i) \sin(\theta).
        \label{eq:vlos_obs}
    \end{split}
\end{equation}
The inclination $i$ of the disk together with the azimuthal angle $\theta$ and the radius $r$ form a polar coordinate frame.

\subsubsection{Adopted tilted ring model for \texorpdfstring{\gal}{M83}}
\label{sec:adopted_tilted_ring_model}
Several kinematic parameters for \gal\ exist in the literature. However, the rotation curve that extends the farthest and is readily available, was published in \cite{Heald2016} using KAT7 (the seven-dish MeerKAT precursor array) observations with an angular resolution ${\sim}$230$\arcsec{\approx}$5.33~kpc. Since \hi\ in \gal\ extends far in our observations, this is the best suited rotation curve that involves the northern and southern arms of \gal\ (see \autoref{sec:Integrated Intensity Map}). In addition, \cite{Heald2016} performed several tilted ring models explicitly for this galaxy, which strengthens our selection of this rotation curve over others. They used the GIPSY task \texttt{rotcur} to construct a rotation curve (\vrot) using a ring width of 100$\arcsec$. For that, they set the radial velocity component (\vrad) to zero which reduces \autoref{eq:vlos_obs} to the first two terms. \cite{Heald2016} found at radii beyond 1000$\arcsec$ ${\sim}$23~kpc an opposite deviation of the modeled rotation curves of both, the approaching and receding side. To account for this, they used \vsys = 510~\kms\ for r$<$23~kpc and \vsys = 500~\kms\ for r$>$23~kpc. In this work we adopt their resulting kinematic parameters from this approach (see also discussion in \autoref{sec:massflow_model}).
    
To get a modeled velocity map (\vlos$_{\mathrm{, mdl}}$) we used their rotation curve (\vrot(r)), their constant inclination angle of 48$\degr$, and, their position angles which vary from 225.0$\degr$ for the central ring to 158.6$\degr$ for the outermost ring (see \autoref{fig:parameter_compare}):
\begin{equation}
        V_{\rm los, mdl}(x, y) = V_{\rm sys} + V_{\rm rot}(r) \sin(i) \cos(\theta).
        \label{eq:vsyn}
\end{equation}
Here, \vsys\ is the systemic velocity of the galaxy with respect to the observer, and \vrot\ the rotation velocity. The \vlosmdl\ is denoted in sky coordinates ($x,y$) while the terms on the right side of \autoref{eq:vsyn} are in the disk coordinate frame ($r, \theta$).  These two systems are related:
\begin{equation}
       \cos(\theta) = \frac{- (x-x_0)  \sin(\mathrm{PA}) + (y-y_0)  \cos(\mathrm{PA})}{r}
       \label{eq:costheta}
\end{equation}
and also $\sin(\theta)$ from \autoref{eq:vlos_obs} as:
\begin{equation}
     \sin(\theta)= \frac{- (x-x_0)  \cos({\rm PA}) - (y-y_0)  \sin(\mathrm{PA})}{r\cos(i)},
    \label{eq:sintheta}
\end{equation}
where $x_0$ and $y_0$ denote the center coordinates and position angle is the angle measured counter-clockwise between the north direction of the sky and the major axis of the receding half of the galaxy. 
We then use \autoref{eq:vsyn} to get \vlosmdl. For the purpose of showing \vlosmdl\ maps (and \vres\ maps, see next paragraph), we interpolate the space between the rings (see \autoref{fig:vres_compare}) and apply a simple mask to only show \vlosmdl\ values that match with our observed S/N masked velocity field (i.e.\ we ignore data outside the mask).

\subsubsection{Residual velocities and radial velocities}    
To determine the residual velocities ($V_{\rm res}$), we subtract \autoref{eq:vsyn} from \ref{eq:vlos_obs}, $ V_{\rm res}(x, y)= V_{\rm los, obs}(x, y) - V_{\rm los, mdl}(x, y)$ and get:
\begin{equation}
        V_{\rm res}(x, y)= V_{\rm rad}(r)\sin(\theta)\sin(i) .
\end{equation}
To get radial velocities (\vrad), we only take \vres\ values within each tilted ring (i.e.\ not using the \vres\ map where we have interpolated between the gaps) and account for the $\sin(\theta)\sin(i)$ term:
\begin{equation}
     V_{\rm rad}(r) = \frac{V_{\rm res}(r)}{\sin(\theta)\sin(i)} . 
     \label{eq:vrad}
\end{equation}
Knowing in which direction a galaxy rotates is required to correctly interpret the nature of measured radial motions. Negative (positive) velocities in \vrad\ are inflow (outflow) motions when a galaxy is rotating clockwise, and outflow (inflow) when a galaxy is rotating counterclockwise. Under the assumption that spiral galaxies spin with their arms trailing the direction of rotation, the winding of the extended \hi\ arms reveal that \gal\ is rotating clockwise. Therefore, \vrad\ $<0$ indicates inflow and \vrad\ $>0$ outflow for \gal.

%
%
\section{Results}
\label{sec:Results}
\begin{figure*}
    \centering
    \includegraphics[width=1.0\textwidth]{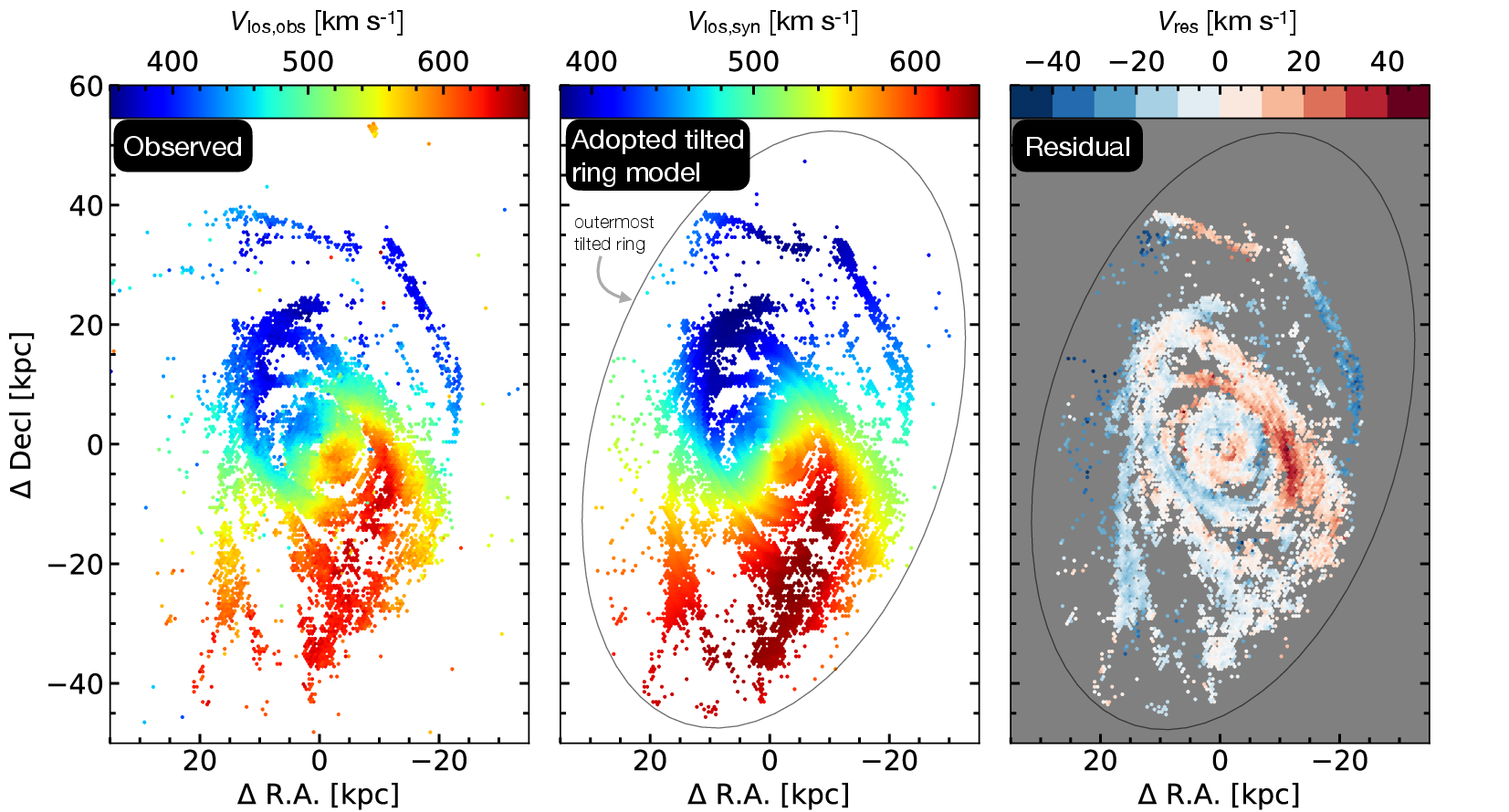}
    \caption{Observed, adopted, and residual velocity maps.\textit{Observed:} Velocity field of \gal\ using our VLA \hi\ observations. The blue colors represent the approaching and the red is the receding side of the \hi\ galaxy disk. \textit{{Adopted tilted ring model:}} Constructed velocity field using \vsys, \vrot, $i$, and position angle of the tilted ring model by \cite{Heald2016}. The green colors represent the systemic velocity of $\sim$510 km~s$^{-1}$ for the inner ${\sim}$23~kpc in galactocentric radius and $\sim$500 km~s$^{-1}$ beyond (see \autoref{sec:adopted_tilted_ring_model}). The tilting results in gaps between the rings, which we interpolated for the presentation of these maps (2d interpolation, see Figure \ref{fig:vres_compare}). We had to restrict our field of view to the model outputs (i.e. the outermost tilted ring). \textit{{Residual: }} The difference between the observed velocities and the modeled velocities; in the range of -40 to 40\kms. We show the uncertainty maps in the Appendix \autoref{appendix:add_figures}}
    \label{fig:mom1}
\end{figure*}

In this section we present the results derived from our \hi\ observations towards \gal\ (integrated intensity map and velocity fields, \autoref{fig:mom0} and \autoref{fig:mom1}), analyzing how the super-extended \hi\ disk of \gal\ compares to its optical central disk. We do this by applying a simple environmental mask by visually separating these regions (see \autoref{fig:diff_reg}) that allows us to distinguish between (i) central disk, (ii) ring, (iii) southern area, (vi) southern arm, and, (v) northern arm. 

\subsection{Distribution of \texorpdfstring{\hi}{HI} in \texorpdfstring{\gal}{M83}} 
\label{sec:Integrated Intensity Map}

In \autoref{fig:mom0} we show the integrated \hi\ intensity map of \gal\ that extends over one degree on the sky. The \hi\ emission towards the central optical disk is ${\sim}8.1$~kpc in radius and shows the highest integrated intensities. The right panel in \autoref{fig:mom0} shows an optical image of M83's central disk with \hi\ column densities in cyan contours. These contours extend up to a radius of 50~kpc (${\sim4}$ times the size of the optical disk).  

The central disk is surrounded by \hi\ emission that follows a ring like structure that extends to a galactocentric radius of ${\sim}16$~kpc. The most prominent features in the outskirts of \gal\ are the southern and northern spiral like structures seen already in previous studies (e.g.\ \citealt{Bigiel2010,Heald2016}). 
Our field of view (grey solid line in \autoref{fig:mom0}) allowed us to also detect the nearby companion galaxy cataloged as UGCA\,365 which is $5.25\pm0.43$~Mpc in distance (TRGB measurements; \citealt{Karachentsev2007}). We measure that this dwarf irregular galaxy is ${\sim}50$~kpc in projected distance from the center of \gal\ and has in our map a diameter of $88\arcsec$($\approx 2$~kpc). This is a factor of two smaller than has been found by \cite{Heald2016}. We find an \hi\ mass of $M_{\rm HI} = 6.1\times10^5~$M$_{\odot}$, which differs from the quoted \hi\ mass in \citealt{Heald2016} who finds $M_{\rm HI} = 2.7\times10^7~$M$_{\odot}$. This is attributed to the fact that our VLA observations are not as sensitive as the KAT-7 data. This companion galaxy was faintly detected in the GBT observations (see \autoref{fig:gbt}). 

Throughout this work we refer to different regions: The \textit{central disk} where we find an averaged \hi\ column density of $6.28\times10^{20}~$cm$^{-2}$, the \textit{ring} that surrounds the central disk, the prominent \textit{northern arm} and \textit{southern arm}. In the southwest we detect significant emission with spots of enhanced column densities -- we refer to this region as the \textit{southern area}.

\subsection{Velocity Fields} 
\label{sec:Velocity and Rotation}
\begin{figure*}
    \centering
    \includegraphics[width=1.0\textwidth]{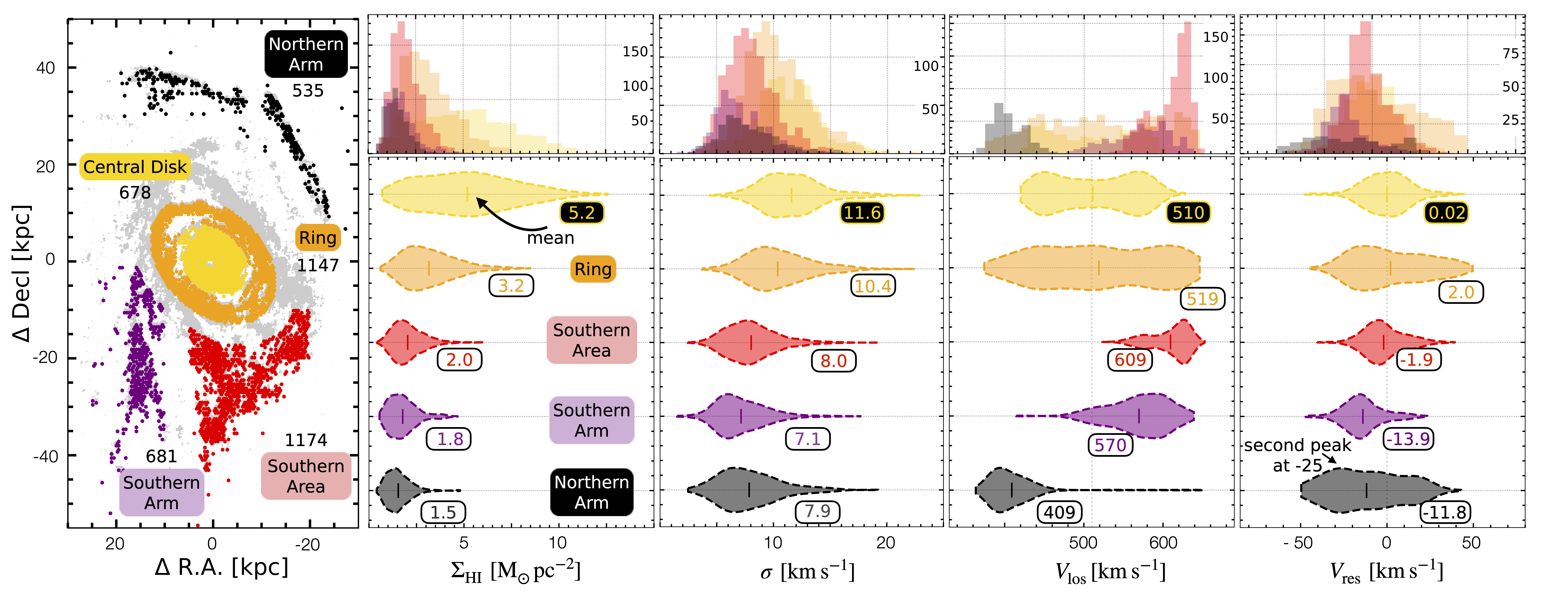}
    \caption{{ Environmental differences in velocities, \hi\ gas surface density and velocity dispersion in \gal.} \textit{Mask:} Colors represent the different regions: Central disk, ring, southern area, northern arm and the southern arm. We note the total sightlines for each of the regions. The grey colors in the background show the \hi\ column density contours ($N_{\rm HI} = 13\times10^{19}$\,cm$^{-2}$) that we showed in \autoref{fig:mom0}.
    \textit{Violin plots:} Each violin represents the distribution of each quantity in a region defined in the mask. We set the kernel density estimation (KDE) to compute an empirical distribution of each quantity to 200 points. The columns on the x axis show the \hi\ gas surface density \sigatom, the second the line width \sigeff, the observed velocities \vlos, and the last residual velocities \vres. The mean value of the observed quantity is reported for each violin. The long tails seen for example in the \vlos\ violin for the northern arm represents that there is one discrepant value.}
    \label{fig:diff_reg}
\end{figure*}
{
\renewcommand{\arraystretch}{1.8}
\begin{table*}
\centering
\caption{Properties of environmental regions in \gal.}
\label{tab:regions}
\begin{tabular}{lccccccc}
\hline \hline
\multicolumn{1}{c}{} & $\langle$ $I_{\rm HI}$ $\rangle$& $\langle$ $N_{\rm HI}$ $\rangle$& $\langle$ \sigatom $\rangle$ & $\langle$ \sigeff $\rangle$ & $\langle$ \vlos $\rangle$ & $\langle$ \vres $\rangle$  & $\langle$ $I_{\rm HI}$/FUV $\rangle$   \\
\multicolumn{1}{c}{\multirow{-2}{*}{Region}} & [K km s$^{-1}$]& [$10^{20}$~cm$^{-2}$] & [M$_{\odot}$ pc$^{-2}$] & [km s$^{-1}$] & [km s$^{-1}$] & [km s$^{-1}$] & [$10^{5}$]   \\
\multicolumn{1}{c}{\multirow{1}{*}{(1)}} & (2) & (3) & (4) & (5) & (6) & (7) & (8)  \\
\hline
central disk & 
$345.2_{161.8}^{514.9}$&
6.3$_{2.9}^{9.4}$&
$5.2_{2.4}^{7.7}$&
$11.6_{9.1}^{13.9}$&
$510.2_{447.6}^{573.6}$&
$0.0_{-12.9}^{12.1}$&
7.3$_{2.0}^{13.1}$ 
\\
ring & 
$210.3_{123.0}^{300.1}$&
3.8$_{2.2}^{5.5}$&$3.2_{1.8}^{4.5}$&
$10.4_{7.8}^{13.1}$&
$519.1_{435.1}^{609.1}$&
$2.0_{-19.9}^{27.5}$&
35.2$_{18.3}^{51.2}$
\\
southern area &
$135.7_{85.6}^{189.0}$&
2.5$_{1.6}^{3.4}$&
$2.0_{1.3}^{2.8}$&
$8.0^{5.9}_{10.1}$ &
$609.4^{578.4}_{631.3}$&
$-1.9^{-12.6}_{10.3}$&
31.8$_{21.5}^{44.2}$
\\
southern arm &
$118.4_{70.8}^{163.8}$&
2.2$_{1.3}^{2.9}$& 
$1.8^{1.1}_{2.5}$&
$7.1^{5.2}_{9.3}$  &
$569.6^{527.2}_{604.9}$&
$-13.9^{-23.9}_{-3.2}$&
27.9$_{17.9}^{38.9}$
\\
northern arm &
$102.7_{61.3}^{141.6}$&
1.9$_{1.1}^{2.6}$&
$1.5^{0.9}_{2.1}$&
$7.9^{5.4}_{10.2}$ &
$408.9^{385.7}_{428.4}$&
$-11.8^{-33.6}_{15.1}$&
25.9$_{15.8}^{36.1}$
\\
\hline 
\end{tabular}
 \begin{minipage}{2.0\columnwidth}
        \vspace{1mm}
        {\bf Notes.} (1): Defined environmental regions in \gal: central disk, ring, southern area, southern and northern arm. (2--8): We show the mean, the 16th, and 84th percentiles of each quantity for each region. (2): \hi\ integrated intensity. (3): \hi\ column density. (4): \hi\ surface density. (5): \hi\ effective line width. (6): L.o.s velocity. (7): Residual velocity. (8): Ratio of \hi\ over the  GALEX FUV map shown in \autoref{fig:fuv_sigma}. This results in units of K km s$^{-1}$/(mJy arcsec$^{-2}$).
    \end{minipage}
\end{table*}}


In this work, we analyze observed and modeled first-moment maps to examine non-circular motions. The observed velocities, \vlosobs\, in the first panel of \autoref{fig:mom1} are ranging from 357 to 661 \kms\, where 510 \kms\ is the systemic velocity (shown as green colors, see also \autoref{Tab: Properties}) for the inner ${\sim}$23~kpc in galactocentric radius and $\sim$500 km~s$^{-1}$ beyond (see \autoref{sec:adopted_tilted_ring_model}). The central disk shows symmetric velocity behaviors that are typical for gas moving in circular motions. However, this symmetry breaks at larger radii. The ring already presents differences in the systemic velocity and the corresponding approaching (blue) and receding (red) velocities; as the symmetry axis is no longer the major and minor axis of the central disk. 
The northern arm only shows approaching velocities, whereas the southern arm shows a transition from systemic to receding velocities along the arm direction. Both of these arms are winding clockwise, similar to the spiral arms seen in the central (optical) disk from, for example, CO observations (e.g.\ \citealt{Koda2020}).

The second panel of \autoref{fig:mom1} shows the line-of-sight velocity field implied by the \cite{Heald2016} titled ring model (we use their \vrot, $i$ and position angle for each tilted ring; we describe the procedure and our motivation to use this set of kinematic parameters in \autoref{eq:vsyn} and \autoref{sec:adopted_tilted_ring_model}). The field of view was restricted to the model outputs, meaning that we neglect everything beyond the outermost tilted ring (including the northern companion galaxy UGCA~365). In this map we see lower approaching velocities towards the beginning of the northern arm and higher receding velocities in the southern area compared to the observed velocities. 

The third panel in \autoref{fig:mom1} shows the residual velocities, \vres. The highest \vres\ values are indicated with dark red or dark blue colors in the third panel in \autoref{fig:mom1}. We find values of $\sim$40 \kms\ in the southwest of the ring along with $\sim -10$ \kms\ on the eastern side. The southern area and southern arm are shown in white to light-blue colors indicating a \vres\ of ${\sim}0$ -- $-10$ \kms. The northern arm shows negative and positive \vres\ values, whereas the kink of the arm (inflection point, see \autoref{sec:UGCA365}) has \vres\ ${\sim}0$ \kms. Additionally, we find at the end of the northern arm a "blob" of negative \vres\ values of  ${\sim} -30$ \kms.

\subsection{Environmental differences in velocities, \texorpdfstring{\hi}{HI} gas surface density and velocity dispersion} 
\label{sec:Environment}


We analyze different environmental regions in more detail shown in \autoref{fig:diff_reg}. We do this by masking each individual area by eye based on the emission seen in \autoref{fig:mom0} after applying a \sn\ cut (\sn\ =3).\footnote{The central disk and the ring are masked by a radius cut (r$_{gal}=8.1$~kpc and r$_{gal}=16$~kpc, respectively) in the plane of the disk. The gap between the central disk and the ring represent lines of sight that were difficult to assign to a specific region and thus were not included in a region. The two arms and the southern area were selected with a cut in R.A. and declination.} This results in the following total number of sightlines separated by a beamsize for each region: for the Central Disk$=678$, Ring$=1147$, Southern Area$=1174$, Southern Arm$=681$, and Northern Arm$=535$. In the following we examine the environmental dependence of \sigatom, \sig, \vlos, and \vres\ in \gal\ shown in \autoref{fig:diff_reg}. 

The second column shows the \hi\ surface density. This quantity is overall decreasing with galactocentric radius, from the central disk to the outermost outskirt region -- the northern arm. We find the highest mean \sigatom\ in the central disk (yellow violin) of $5.2~$\sigatomunit. The ring (orange violin), southern area (red violin), and southern arm (purple violin) show mean \sigatom\ of $3.2~$\sigatomunit\ , $2.0~$\sigatomunit\ and $1.8~$\sigatomunit\, respectively. The northern arm (black violin) exhibits a mean \sigatom\ of $1.5~$\sigatomunit\; more than a factor of ${\sim}3$ lower then the central disk. The overall trend of decreasing \sigatom\ with $r_{\rm gal}$ in \gal\ agrees with previous studies (e.g.\ \citealt{Bigiel2010}).

The third column shows to first order that the means of the velocity dispersion (using \sigeff, the effective width; see \autoref{eq:sigma}) decrease with $r_{\rm gal}$. Upon closer examination, it becomes clear that the northern arm has a slightly higher mean line width than the southern arm. However, the overall distribution of those two regions look reasonably similar. Higher values in \sigeff\ could be attributed to multiple components and/or wider components. This way of measuring the velocity dispersion does not distinguish between these possibilities (see Discussion in \autoref{sec:velocitydispersion}). The overall trend of decreasing \sig\ with $r_{\rm gal}$ agrees with previous studies (e.g.\ \citealt{Tamburro2009} and see our discussion in \autoref{sec:velocitydispersion}). 

The last two columns show the line-of-sight and residual velocities (i.e.\ \vlos\ and \vres). In general, the central disk is well described by the titled ring model (that assumes circular motions) as the distribution of \vlos\ looks very symmetric and the mean of \vres\ is close to zero ($\langle$\vlos$\rangle$ $=0.02\,$\kms). The mean value of $510$~\kms\ is the same as used for systemic velocity in the tilted ring model ( until a radius of $\sim$23~kpc; see \autoref{sec:adopted_tilted_ring_model}). The ring region is still relatively well-characterized by this model, with deviations from pure circular motions of $\langle$\vres$\rangle =2.0\,$\kms. Also, the southern arm has low mean deviations of $\langle$\vres$\rangle =-1.9\,$\kms, however, the mean \vlos\ values are on the receding side (i.e.\ $>510~$\kms, grey dashed horizontal line). The outermost regions, the southern and northern arms, show significant deviations from circular motions. They exhibit $\langle$\vres$\rangle$ values of $-13.9\,$\kms\ and $-11.8\,$\kms, but their means in \vlos\ are on the receding and approaching side, respectively. The distribution of \vres\ values towards the northern arm represents a change in velocities along the arm (we discuss that in \autoref{sec:UGCA365}). 

We show for each quantity the mean, the 16th and 84th percentiles in \autoref{tab:regions}. This table already contains values of the I$_{\rm HI}$/FUV ratio for each region. We discuss a visual comparison of FUV emission (as a tracer of star formation rate) and velocity dispersion later in the paper (see \autoref{sec:velocitydispersion}). The table reveals that the central disk has the smallest I$_{\rm HI}$/FUV values (i.e.\ $7.32\times10^{5}~$ K km s$^{-1}$/(mJy arcsec$^{-2}$)). The ring shows the largest values and I$_{\rm HI}$/FUV then decreases with $r_{\rm gal}$.

%
%
\section{Discussion} 
\label{sec:Discussion}
In this section, we study the radial dependence of the velocity dispersion (obtained with three different methods) and \hi\ surface density and how it changes beyond the central disk. We further examine possible reasons for broader \hi\ profiles and discuss the extended \hi\ structure of \gal\ and its companion UGCA~365. 

\subsection{\texorpdfstring{\hi}{HI} Velocity Dispersion in the outskirts of \texorpdfstring{\gal}{M83}}
\label{sec:velocitydispersion}

\begin{figure}
    \centering
    \includegraphics[width=0.48\textwidth]{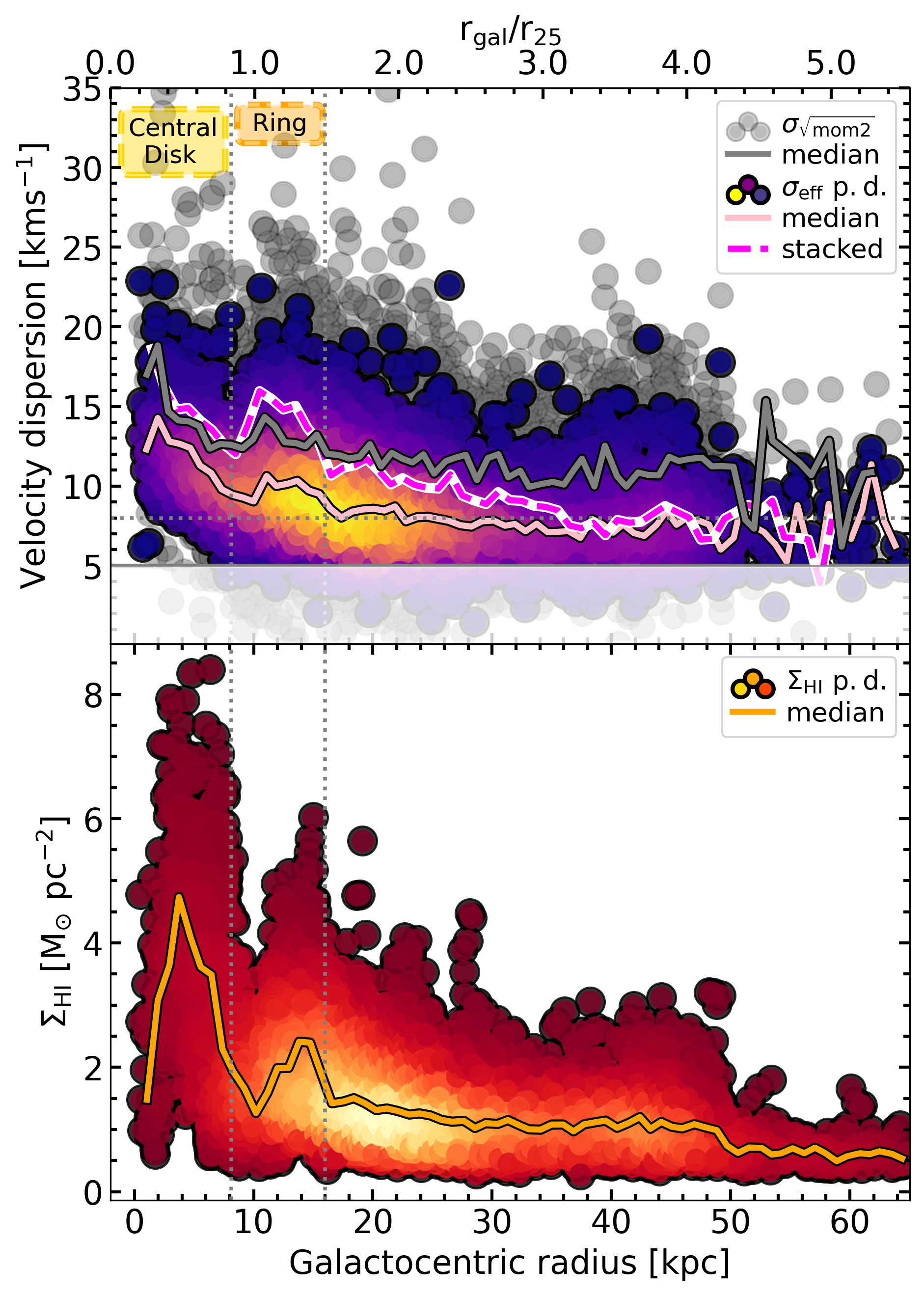}
    \caption{Radial profile of the velocity dispersion, $\sigma_v$, and the \hi\ gas surface density \sigatom, together with the point density of $\sigma_v$ and \sigatom\ for individual lines of sight (circular points; the color-coding reflects a linear density distribution, where the highest density is shown in yellow).  \textit{Top panel}: We compare here the effective width (\sigeff\, purple-to-yellow scatter points) and $\sqrt{\rm mom2}$ (\sigmom\, grey points) approaches to estimate the velocity dispersion across the whole disk of \gal. We show the median for \sigmom\ (grey solid line), the median for \sigeff\ (pink solid line), and the results after stacking \sigeff\ (pink-white dashed line). We see in all three cases enhanced velocity dispersion in the central disk and in the ring (indicated by the dashed vertical lines at 8.1~kpc and 16~kpc). The horizontal dashed line indicates the 8 \kms\ limit that we discuss in \autoref{sec:velocitydispersion}. \textit{Bottom panel:} The profile of the \hi\ gas surface density shows an enhancement of \hi\ gas in the ring. This is similar to the velocity dispersion profiles. }
    \label{fig:sig_rad}
\end{figure}
Multiple observations have shown that the \hi\ velocity dispersion encodes information on a combination of temperature, turbulence, and unresolved bulk motions and is thus sensitive to feedback and energetics and the physical state of the gas \citep[e.g.][]{Tamburro2009,Ianjamasimanana2015,Mogotsi2016,Romeo2017,Koch2018,Ianjamasimanana2015,Oh2022}. In the upper panel of \autoref{fig:sig_rad} we show the radial profile of the velocity dispersion across the whole disk of \gal. Overall the velocity dispersion decreases with galactocentric radius, similar to the trend we found in \autoref{fig:diff_reg}. This agrees also with studies where it was shown that $\sigma_v$ decreases with $r_{\rm gal}$ across local disk galaxies (e.g.\ \citealt{Tamburro2009}). However, in our radial profile, a peak of velocity dispersion is evident in the region of the ring, which we marked with the two vertical dashed lines. 

We compare here three radial profiles obtained with different methods that are commonly used in the literature. We first compare the effective width approach (\sigeff\ colorized markers; see \autoref{eq:sigma}) with the square-root of the second moment (\sigmom\ grey markers; see \autoref{eq:sigma_mom2}) and show their corresponding profile (grey and pink solid lines). \sigeff is used to study cooling and the physics of the ISM, on the other hand, for studies dealing with dynamics and large-scale evolution, using \sigmom\ is more appropriate. We additionally show the velocity dispersion profile where we stacked the \sigeff\ values (pink-whited dashed line; see \autoref{sec:binning_stacking} for a description of the binning and stacking techniques). We find that the profiles for \sigmom, \sigeff, and stacking \sigeff\ all show, in addition to the center, enhanced velocity dispersion in the ring. The same behavior arises in the \sigatom\ radial profile shown in the bottom panel. 

From \autoref{fig:sig_rad} it is immediately apparent that the profile of \sigmom\ has a higher velocity dispersion than that of \sigeff. However, the stacked profile shows an alternative behavior; being higher than the median values until r$\sim$40~kpc and then approaching the lower values of the median. Overall, the different approaches show similar trends, but are offset by ${\sim}3\,$\kms\ between \sigeff\ and \sigmom, and $2\,$\kms\ between \sigeff\ and stacked \sigeff\ (taking the mean between the differences until $r_{\rm gal}=40$~kpc). The study by \cite{Koch2018} analyzed the differences between these methods in more detail. They demonstrate that the deviation in \sigmom\ to the stacked profiles are strong indications for multiple velocity components in some lines of sight.
An analysis towards M\,31 and M\,33 (\citealt{Koch2021}) finds $>$50$\%$ lines of sight have more than one Gaussian in a spectrum, suggesting that multi-Gaussian profiles can be a source of discrepancy. This has also been found in various other nearby galaxy surveys (\citealt{Warren2012,Stilp2013a,Stilp2013b}). An examination of the masked ring region reveals some of these features (see \autoref{appendix:add_figures}). A by-eye estimate, suggests more than one third of the $>$1000 lines of sight in the ring with $>$1 components. A more sophisticated fitting of individual spectra is beyond the scope of this paper and is limited by our need to increase the \sn\ via smoothing the data to 5 \kms\ channels (see \autoref{sec:imaging}). Nevertheless, \sigeff\ and \sigmom\ show an enhancement in velocity dispersion in the ring.

Line widths of ${\sim}6{-}10$~\kms\ have been observed in the outskirts of galaxies (e.g.\ \citealt{Tamburro2009} using the second-moment approach). At pressures that are typical for disk galaxies, the interstellar atomic gas has two phases at which it can remain in stable thermal equilibrium \citep[e.g.][]{Field1969,Wolfire1995,Wolfire2003}. These two phases correspond to the cold (CNM) and warm neutral media (WNM) with temperatures of ${\sim}100$~K and ${\sim}8000$~K, respectively. The cold neutral gas emits lines with a characteristic line width of ${\sim}1$~\kms\ , while the warm neutral gas has a line width of ${\sim}8$~\kms\ . \cite{Tamburro2009} found spectral lines broader than ${\sim}8$~\kms\ in the THINGS galaxies and interpreted them to be broadened due to turbulent motions. However, as mentioned in the previous paragraph, the second-moment approach has been found to overestimate line width \citep[e.g.][]{Mogotsi2016,Koch2018}. 

The bulk of the gas in the outskirts of galaxies is expected to be in the form of WNM (see e.g.\ \citealt{Dickey2009} , who find a CNM fraction of ${\sim}$15-20\% out of $r{\sim}$25~kpc in the Milky Way). The WNM temperature does not drop below ${\sim}$5000 K (corresponding to \sig\ ${\sim}$6~\kms\ ) and is typically closer to $6000{-}7000$ K (see e.g.\ \citealt{Wolfire2003}). Accordingly, the WNM is not expected to have a thermal velocity dispersion larger than 8~\kms. We find that the velocity dispersion is higher than the thermal WNM line width (i.e., greater than 8 \kms; marked as the horizontal dashed line in \autoref{fig:sig_rad}) for both \sigmom\ and \sigeff. This agrees with past studies \citep[e.g.][]{Tamburro2009,Koch2018,Utomo2019}. Our observations, however, highlight the large galactocentric radius range where we find a shallow decrease in velocity dispersion that is slightly larger than thermal -- until $r_{\rm gal}{\sim}$50~kpc. The exact values of the velocity dispersion shown in \autoref{fig:sig_rad} should be taken with caution, since individual spectra in our observations show multiple peaks (especially in the central disk) for which associating those values with a purely thermal line width may not be appropriate.

\begin{figure}
    \includegraphics[width=0.49\textwidth]{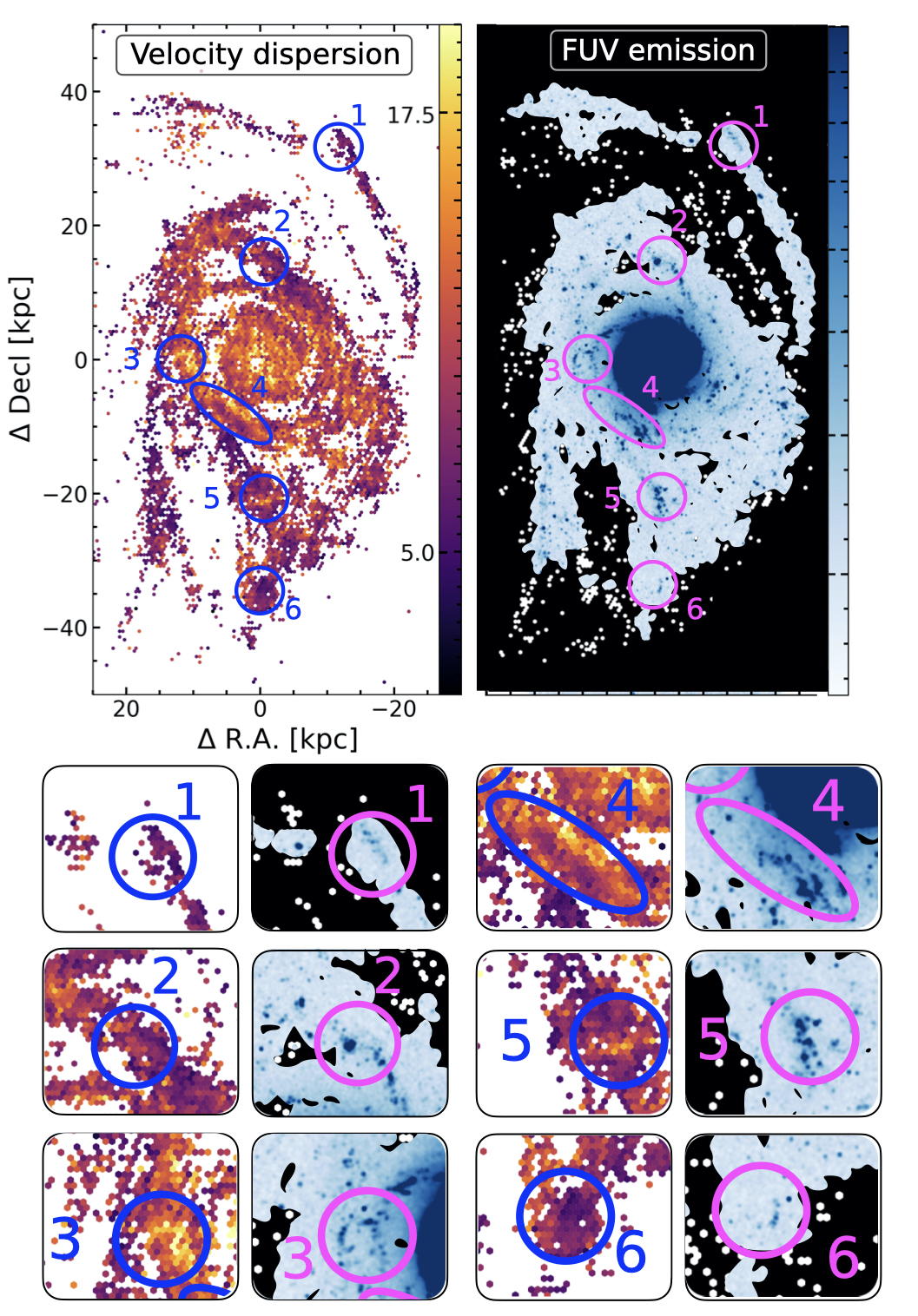}
    \caption{Velocity dispersion and FUV emission. Visual inspection of spatial correlation of velocity dispersion (\sigeff\, the colorbar represents units of \kms) and FUV emission as an indicator of star formation (in units of mJy arcsec$^{-2}$). The FUV emission is the same GALEX FUV map as shown by \cite{Bigiel2010}. We defined 6 individual sub-regions in the ring, southern area, southern and northern arm, based on where we see either stronger FUV emission or enhanced velocity dispersion. See \autoref{sec:velocitydispersion} for the discussion on the different regions. }
    \label{fig:fuv_sigma}
\end{figure}

We investigate whether there is evidence for an enhancement in velocity dispersion that is connected to ongoing star formation as would be expected if turbulent motions were driven by feedback. \gal\ has an extended UV (XUV) disk (\citealt{Thilker2005}) where a tight spatial correlation between \hi\ and FUV emission has been found (\citealt{Bigiel2010}). Using their FUV map, we see at first sight within the ring peaks of FUV emission. However, a visual inspection of the spatial correlation of velocity dispersion and FUV emission (as an indicator of star formation) towards sub-regions in the ring shows at the location with the highest FUV emission the lowest \hi\ velocity dispersion (e.g. region 4 in \autoref{fig:fuv_sigma}). Furthermore, regions labeled as 1, 2, 4, 5 and 6 show peaks in their FUV emission but no strong velocity dispersion. Region 3 shows within their circular regions stronger FUV emission together with enhanced velocity dispersion, however, they do not perfectly spatially correlate. That indicates that at least at face value there is no immediate one-to-one correspondence between recent star formation and enhanced velocity dispersion. 

Whether the cause of higher velocity dispersion is feedback (from star formation) or gravity-driven (gravitational instabilities) turbulence is still a matter of debate in the literature (see e.g. discussion by \citealt{Krumholz2016}). Some authors found that high velocity dispersion is well correlated with regions of active star formation. However, for example, \cite{Krumholz2016} argue that the correlation itself may just be explained by the fact that galaxies with more gas tend to have both higher velocity dispersion and more star formation. In a model combining both star formation feedback and radial transport, \cite{Krumholz2018} have shown that turbulence in galaxy disks can be driven by star formation feedback, radial transport, or a combination of both. Our observational analysis of \gal\ shows no clear correlation 
with star formation, suggesting that radial transport (e.g. \citealt{Schmidt2016}) could be the reason for enhanced velocity dispersion in the outskirts of \gal.

\subsection{The extended \texorpdfstring{\hi}{HI} structure and the companion UGCA~365}
\label{sec:UGCA365}
\begin{figure}
    \centering
    \includegraphics[width=0.49\textwidth]{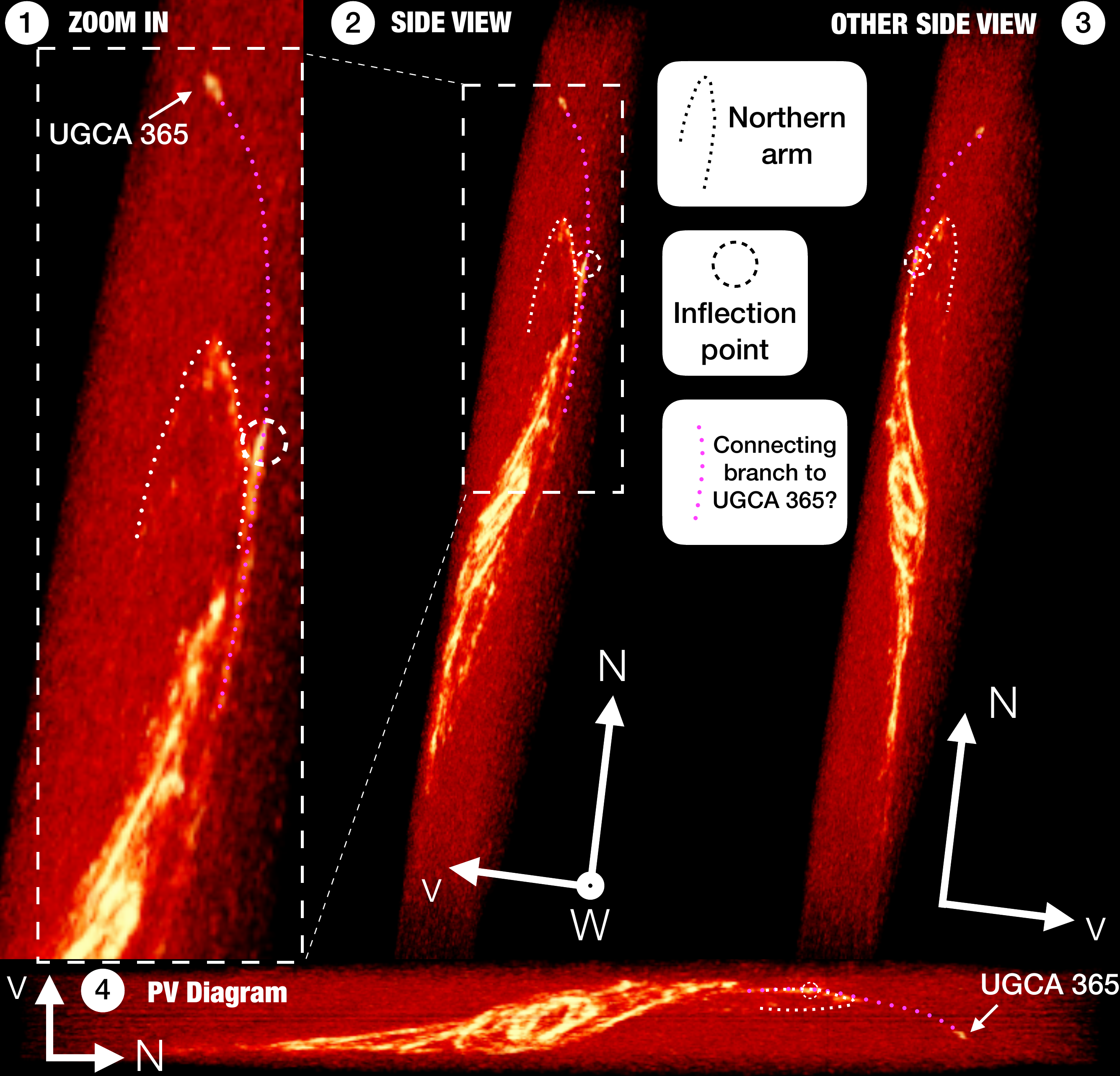}
    \caption{Intensity map of \gal\ from different angles showing the northern arm in a 3D viewer tool (see \autoref{sec:UGCA365}). In the two left plots, the arrow pointing west is directed towards us. In the right plot, it is pointing away from us. The northern orientation is the same in both plots. The v direction indicates the velocity axis over which we integrate for the integrated intensity map shown in \autoref{fig:mom0}. The white doted line shows the direction of the northern arm, while the pink dotted line shows the ${\sim}$20~kpc long speculative connecting branch to UGCA~365.}
    \label{fig:northernarm}
\end{figure}

The sharp edges to the \hi\ gas distribution, especially in the southern and northern part of the extended \hi\ disk, were already noticed by \citealt{Heald2016} (see their figure 17). They consider that this is due to photoionization or ram pressure from the intergalactic medium and specifically ruled out technical issues. 

The prominent northern extended arm emerges from the western part of \gal\ that curves 180$\degree$ around to the east. This can be best seen in the observations of \cite{Heald2016} and \cite{Koribalski2018} serving as evidence that \gal\ may have interacted or merged with another, smaller galaxy. In our observations the eastern part of the extended arm is not detected and the irregular galaxy NGC\,5264 is not in our field of view (see \autoref{fig:pointing} and \autoref{fig:gbt}). \cite{Koribalski2018} mentioned a clump marking a kink in the north-west side of the arm. We also see that and refer to it as the "inflection point" because it is also visible in the velocity and residual maps where the velocities change\footnote{The exact velocity values at the inflection point depend on the assumed orientation and the \vrot\ values. For this reason, velocity gradients are not discussed.} and we see enhanced FUV emission at this location (see region 1 in \autoref{fig:fuv_sigma}). We marked this inflection point in \autoref{fig:northernarm} together with the northern arm. In the enclosed snapshot using a 3D viewer tool \footnote{Glnemo2; \url{https://projets.lam.fr/projects/glnemo2/wiki/}, \cite{Lambert2012}}, we can see the dwarf irregular galaxy UGCA\,365 in the northernmost region of our field of view; ${\sim}55$~kpc north of the center of \gal. Its TRGB distance is similar to that of \gal\ within the given uncertainty; 5.25$\pm$0.42~Mpc (\citealt{Karachentsev2007}). \cite{Koribalski2018} noted that both, the stellar and the gas distribution show some extension to the southeast along the minor axis of the galaxy and suggest that this may be due to tidal interaction with \gal. In \autoref{fig:northernarm} we denote a ${\sim}$20~kpc long speculative connecting branch between the inflection point and UGCA\,365 (pink dotted line). We detect faint \hi\ emission close to the inflection point in the direction of this branch. The inflection point is best seen in panel 2 and in the zoomed-in version, where it appears to deviate from the general direction of the northern arm (white dotted line). In general, these panels show the warped nature of \gal's super-extended \hi\ disk.

%
%
\section{Upper limits for average radial mass flow rates in M83}
\label{sec:massflow}
Mass flows -- in- and outflow -- play an important role in the process of the evolution of a galaxy; fueling, for example, the inner central molecular zones with fresh new star-forming material (e.g. \citealt{Kormendy2004}, or \citealt{Henshaw2022} for the CMZ of the Milky Way). In this section we discuss mass flow rates, their limitations and how they depend on initial disk parameters. 
%
\subsection{Mass flow profile -- a simple view}
\label{sec:massflow_model}
\begin{figure}
    \centering
    \includegraphics[width=0.49\textwidth]{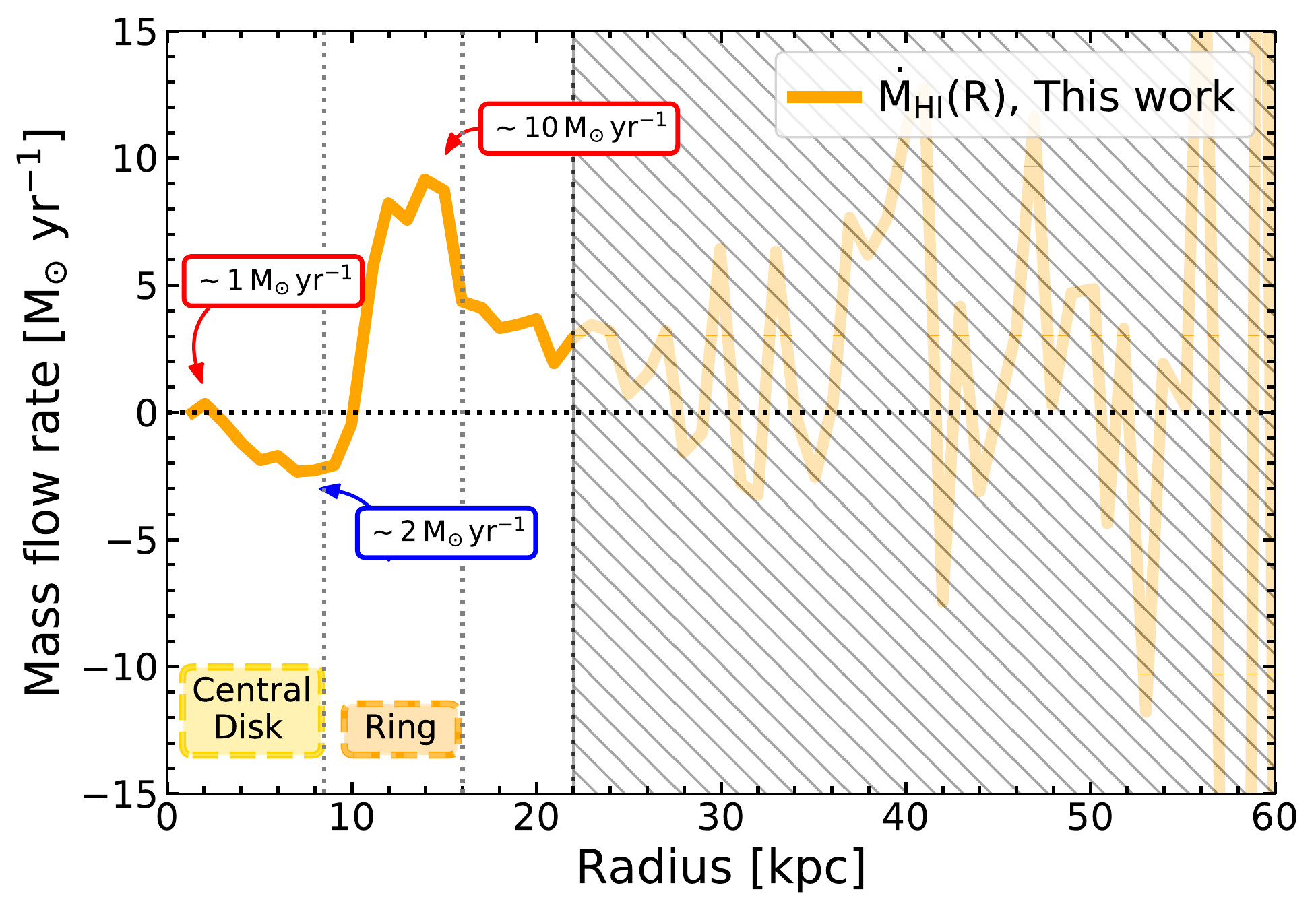}
    \caption{Mass flow rates. Negative (positive) values mean inflow (outflow). We indicate the vertical line where the symmetry in our observed velocities is no longer valid and thus inferred the mass flow rates (beyond galactocentric radius of 23~kpc). We denote the approximate mass flow rates. The peaks and dips are indications in which radial bin in \gal\ gas moves to the previous or following radial bin.}
    \label{fig:massflow}
\end{figure}
\begin{figure*}
    \centering
    \includegraphics[width=1.0\textwidth]{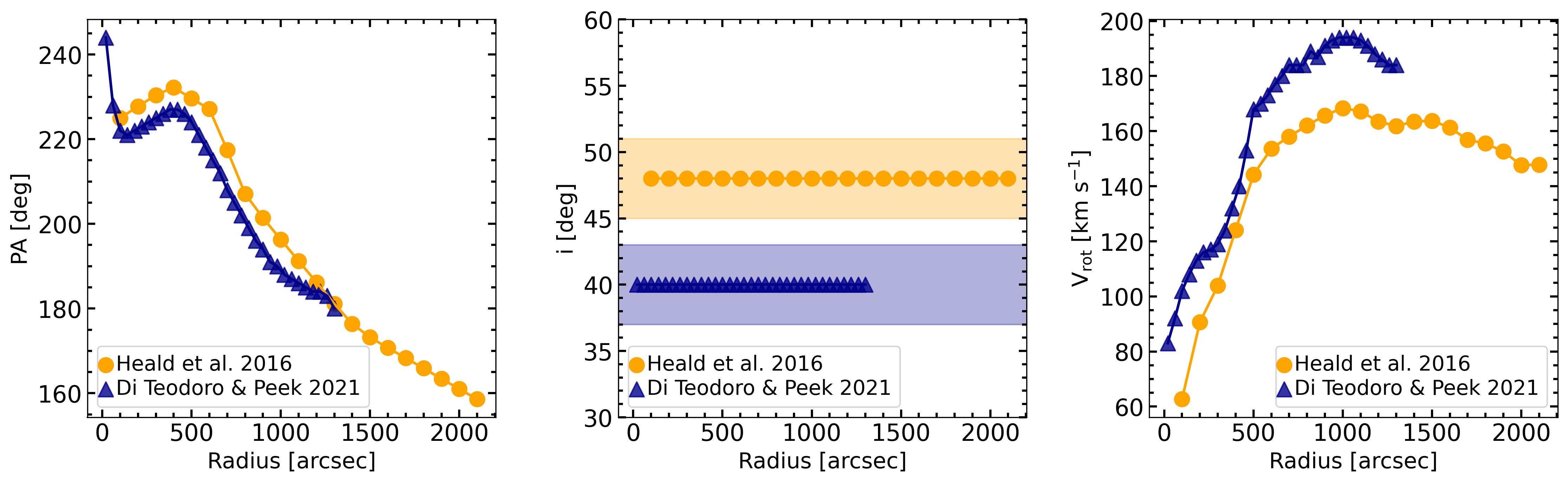}
    \caption{Parameter comparison. In orange we show the kinematic parameters we used in this work for the velocity field shown in \autoref{sec:Velocity and Rotation}. In \autoref{sec:limitations} we investigate how different initial parameters affect mass flow rates using the approach from \cite{DiTeodoro2021} using their own set of parameters for \gal\ (blue markers). }
    \label{fig:parameter_compare}
\end{figure*}

For average mass flow rate profiles across the \hi\ disk in \gal, we use a simplified approach presented for example in \citealt{DiTeodoro2021}: 

\begin{equation}
     \frac{\dot{M}_\mathrm{HI}(r)}{\mathrm{{[M_{\odot}~yr^{-1}]}}} = \frac{2\pi~r}{\mathrm{[pc]}} \frac{\Sigma_
     \mathrm{HI}(r)}{\mathrm{[M_{\odot}~pc^{-2}]}}~\frac{V_\mathrm{rad} (r)}{\mathrm{[pc~yr^{-1}]}},
     \label{eq:mass_flow}
    \end{equation}
where the radial velocity profile \vrad($r$) and \hi\ surface density profile $\Sigma_\mathrm{HI}$($r$) are used to obtain \hi\ mass flow rate profiles. To get radial velocities, \vrad, we use \autoref{eq:vrad}. We remind the reader that \vrad\ $<0$ indicates inflow and \vrad\ $>0$ outflow for \gal\ (see \autoref{sec:Data_products}).

We show the mass flow rate profile in \autoref{fig:massflow}, and also note the radial range of the central disk and ring. \citealp{Heald2016} noticed that beyond a radius of ${\sim}$22~kpc (they quote 1000$\arcsec$) the approaching and receding side rotation curves differ, and thus, should not be over-interpreted. Therefore, we shade the region where the gas distribution and kinematics are no longer symmetric (i.e.\ deviation from pure circular motions).

Based on \autoref{eq:mass_flow} we find within the central disk: (i) at radii ${\sim}2$~kpc evidence of outflowing-material with mass flow rates of ${\sim}$1~M$_{\odot}$~yr$^{-1}$ and (ii) at r${\sim}5.5$~kpc inflowing-material of ${\sim}$2~M$_{\odot}$~yr$^{-1}$. We find indications for (iii) out-flowing material of order ${\sim}$10~M$_{\odot}$~yr$^{-1}$ at radii of $r{\sim}14$~kpc, that is the ring (see mask in \autoref{fig:diff_reg}). In this region we also find higher residual velocities. Moreover, using the \vlosobs\ map we find enhanced velocity dispersion compared to the edges of the central disk of ${\sim}$20\,\kms\ in that same ring region (median value; \sigeff\ and \sigmom, see \autoref{fig:sig_rad}).

\begin{figure*}
    \centering
    \includegraphics[width=0.9\textwidth]{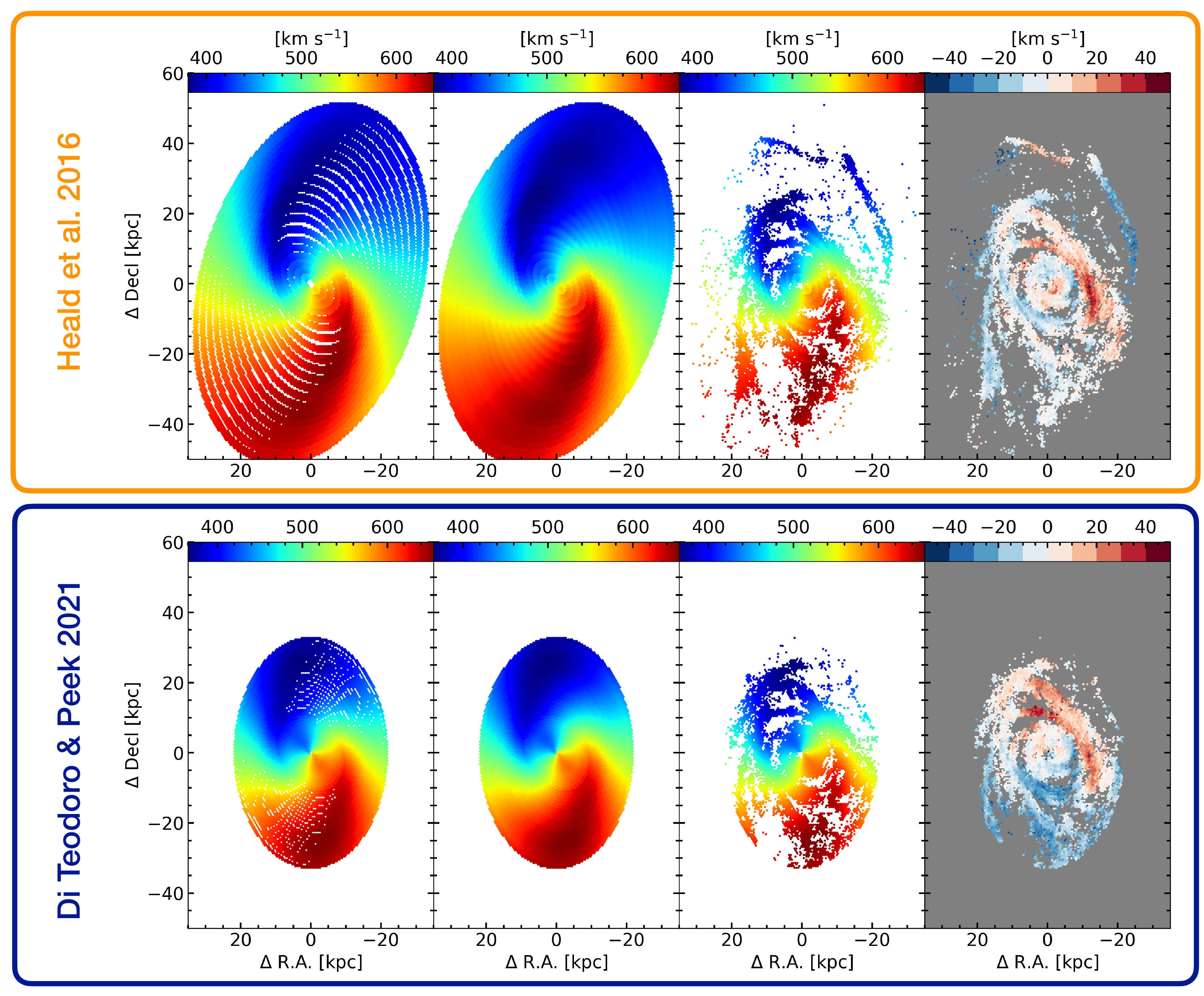}
    \caption{{Comparison of residual maps.} Here we compare the modeled and residual maps (\vlosmdl\ and \vres) of different initial parameters (shown in \autoref{fig:parameter_compare}) that we obtained by using \autoref{eq:vsyn}. \textit{Upper row:} The tilted rings using \citep{Heald2016} model outputs that are based on their KAT7 observations and using \texttt{rotcurv}. For visualisation reasons, we interpolated the gaps using the \texttt{scipy} function 2d interpolation. For further analysis we focus on the tilted-ring frame. \textit{Lower row:} The tilted rings using \citep{DiTeodoro2021} model outputs that are based on LVHIS observations and using \texttt{BBarolo}. The field of view is restricted to their \vrot\ parameters. Therefore the southern and northern arm are not evident in the residual map.}
    \label{fig:vres_compare}
\end{figure*}
\begin{figure}[ht!]
    \centering
    \includegraphics[width=0.45\textwidth]{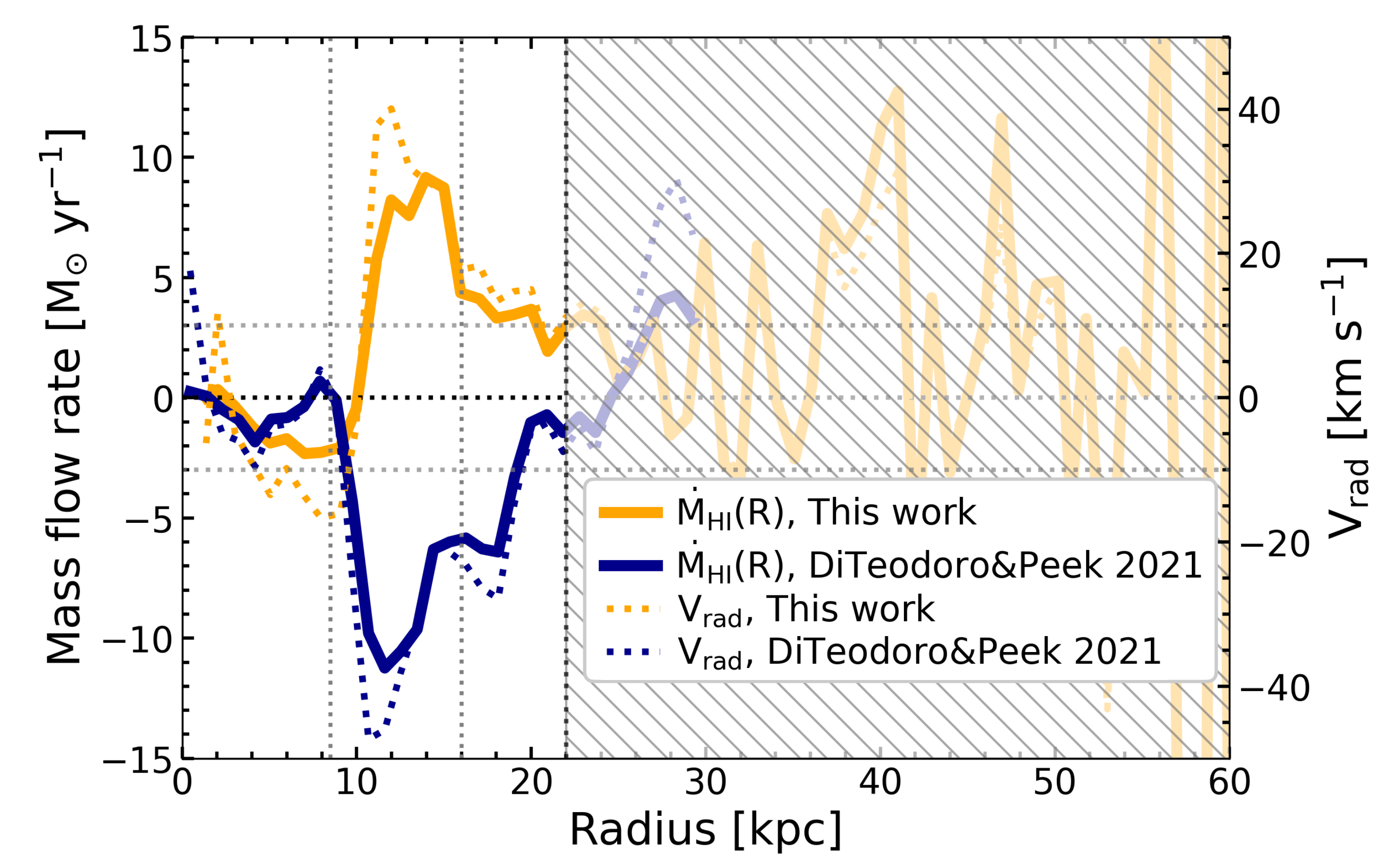}
    \caption{We show here additionally to the mass flow rate of this work (orange) the mass flow rate using different initial parameters as shown in \autoref{fig:parameter_compare} (blue). The dashed lines show \vrad. Negative (positive) values mean inflow (outflow). Within the ring region we find the opposite trend of mass flow rates. 
    We discuss the reasons for this in \autoref{sec:limitations}.}
    \label{fig:massflow_comp}
\end{figure}

\subsection{Initial parameters and their impact on ring-averaged mass flow rates }
\label{sec:limitations}

Multiple variants of tilted-ring codes exist in the literature, for example, the 2D fitter GIPSY task \texttt{rotcur} (\citealt{Begeman1989}), or the 3D fitter \texttt{BBarolo} (\citealt{DiTeodoro2015}),  and \texttt{FAT} (\citealt{Kamphuis2015}). The next step is then often to interpret the non-circular motions and relate them, for example, to flow motions and/or interactions with a companion galaxy. \gal\ is one of the best examples of different kinematic parameters that are available in the literature.   

In \autoref{fig:parameter_compare} we show another recently published  set of kinematic parameters of \gal\ -- position angle, inclination, \vrot\ and \vrad\ with observations from LVHIS using \texttt{BBarolo} (by \citealt{DiTeodoro2021}; extracted from their Appendix Figure 15 for \gal\, also known as NGC~5236). That is the same work from where we adopted the simplified method to derive mass flow rates (see \autoref{eq:mass_flow}). Their tilted rings, however, only extend to $r_{\rm gal}=30$~kpc and therefore not including the southern and northern extended arms. We show the resulting residual velocity maps of both sets of kinematic parameters and how we interpolated the tilted rings (only for visual purposes) in \autoref{fig:vres_compare}. 
Comparing the two \vres\ maps, we see no great difference in the inner disk. In the residual map of \cite{DiTeodoro2021} the deviations of circular motions in the ring region are smaller than in the one of \cite{Heald2016}. In the southern area we see the opposite behaviour; higher deviations of circular motions in the \vres\ map of \cite{DiTeodoro2021}. We take their kinematic parameters shown in \autoref{fig:parameter_compare} and fit them with the same procedure as we did with the kinematic parameters from \cite{Heald2016} (see \autoref{sec:Data_products}). After that we calculate again the mass flow rate using \autoref{eq:mass_flow}. 

We show the two average radial mass flow rate profiles in \autoref{fig:massflow_comp}. What is immediately noticeable is that the two are drastically different. In the ring region, they even indicate the opposite trend -- in-flowing material. \cite{DiTeodoro2021} found a lower inclination of $i=40\degree$ compared to $i=48\degree$. The differences in position angle shift the minor axis, and together with the lower inclination lead to a flip in the inferred flow direction; from outflow to inflow. Additionally, also using LVHIS observations,\cite{Kamphuis2015} derived kinematic parameters using 3D fitting \texttt{FAT} and get an inclination value of $i=40.3\degree$ \footnote{We do not show the calculated mass flow rates for the inclination value given in \cite{Kamphuis2015} as they do not provide any table or radial plots for their position angles where we would be able to extract them.} for \gal. In other words, \cite{Kamphuis2015} and \cite{DiTeodoro2021} find similar, lower values for the inclination than \cite{Heald2016}. 

However, \cite{Heald2016} also determined kinematic parameters using the THINGS data (presented by \citealt{deBlok2008}) and obtained higher values for the inclination, $i=46\degree$. For this purpose, they use the 2D fitting routine \texttt{rotcur}. This higher inclination value and the one derive from KAT7 observations (i.e.\ $i=46\degree$ and $i=48\degree$, respectively), better agree with isophote orientations in the optical disk. 

This is a good demonstration of how challenging it is to unambiguously determine inflow/outflow rates. The inclination depends on whether the kinematics are fitted with 2D or 3D tilted rings (\texttt{BBarolo} tends to fit 10$\degree$ lower inclinations than 2D fitters; \citealt{DiTeodoro2015}). Even at inclinations far from face on, the chosen inclination can have a large impact on the inferred mass flow rate. Mass flow rates are sensitive to the inclination and position angle, to the extent that inflows can even flip sign. At the same point, mass flow rates seem not to be sensitive to the used 3D fit (adopted by \citealt{DiTeodoro2021}) or 2D fit (used here). Also, "circular motion + axisymmetric radial motion" (see \autoref{eq:vlos_obs}) is not guaranteed to yield the true inflow rate, if the \vrot\ $+$ \vrad\ model is not accurate. For these reasons, the values given in \autoref{fig:massflow} and \autoref{fig:massflow_comp} should be considered with caution, though we point out that they do represent our best guess at the \hi\ mass flow rates and directions in \gal. 

In the velocity fields (\autoref{fig:mom1}) and in \autoref{fig:northernarm} we saw the warped nature of M83's super-extended \hi\ disk. The position angle twists by almost 90$^\degree$ from the orientation of the central disk. As mentioned by \cite{Heald2016} this shift in isovelocity was accounted by letting the $v_{\rm sys}$ parameter change in their models (which we adopted in this work, see \autoref{sec:adopted_tilted_ring_model}). However, it has been shown that bar streaming motions are hard to neglect when analyzing mass flow rates towards Milky Way's central molecular zone \citep[e.g.][]{Kim2012,Sormani2015c,Sormani2019c,Tress2020a}, or in nearby galaxies \citep[e.g.][]{Erroz-Ferrer2015,Salak2019}. \gal\ is cataloged as SAB(s)c (see \autoref{Tab: Properties}) -- in between the classifications of a barred spiral galaxy and an unbarred spiral galaxy. The bar within M83's central disk is the main reason for gas inflow seen at $r{\sim}5.5$~kpc, thus the inferred inflow rates at these radii might be underestimated. However, both mass flow rate profiles shown in \autoref{fig:massflow_comp} are consistent for these small radii that cover the bar.

\subsection{Interpreting mass flow rates in \texorpdfstring{\gal}{M83} }
\label{sec:interpretingmassflowrates}

Keeping the uncertainties in mind, the following discussion is based on an approximation of upper limits of mass flow rates in \gal. We compare our results (presented in \autoref{fig:massflow}) to multi-wavelength literature studies (i.e.\ further evidence for mass flows) to interpret our findings.

Starting from the very center of \gal, a study identified signatures of inflowing gas towards its inner circumnuclear ring (${\sim}$130~pc in radius) using ALMA HCN(1-0) and HCO$^{+}$(1-0) observations (\citealt{Callanan2021}). Recently published results towards \gal's optical disk using MUSE observations have shown indications of outflowing ionized gas (\citealt{DellaBruna2022}). This flow is east of \gal's nucleus and ${\sim}1.2$~kpc\footnote{We use here and in the following a distance of 5.16~Mpc, the same mentioned in \autoref{Tab: Properties}, to get physical scales.} in size. This could be consistent with our \hi\ observations indicating an outflow at galactocentric radius of ${\sim}$2~kpc. 

The kinematics over the whole central optical disk of \gal\ has been studied, for example, in \cite{Lundgren2004} and \cite{Fathi2008}. \cite{Lundgren2004} found, using CO(2--1) and CO(1--0) observations, molecular streaming motions along the spiral arms; starting at $r_{\rm gal}{\sim}5.6$~kpc\footnote{See their Figure 8 with the maximal radial extent of 4$\arcmin$.} towards \gal's center. That agrees with our potential (ii) inflow at $r{\sim}5.5$~kpc. \cite{Fathi2008} confirmed this spiral inflow using Fabry-Perot observations of the H${\alpha}$ line across the inner ${\sim}2.3$~kpc of the central disk, capturing this inflow down to a few tens of pc from the dynamical nucleus of \gal. For the possible out-flowing material in the outermost region at radii of $r_{\rm gal}{\sim}14$~kpc we find no correspondence in the literature.

From a theoretical point of view, gas is most likely first accreting onto galaxies at larger radii and then moving radially towards the inner disk to feed star formation (e.g.\ \citealt{Ho2019}). Cosmological hydrodynamical simulations of Milky-Way-like galaxies found expected gas inflow of a few \kms\ within the inner optical disk that reaches ${\sim}$10~\kms\ in the outskirts (e.g.\ \citealt{Trapp2022}). They also found that the gas accumulates at the disk edge and then decreases in average radial speed and increases in column density. Observations also show similar inflow velocities of a few \kms. For example, \cite{Wong2004} decomposed \hi\ and CO velocity fields of concentric elliptical rings into a third-order Fourier series for seven nearby spiral galaxies. \cite{Schmidt2016} performed a Fourier decomposition of velocity fields for 10 THINGS galaxies. Both of these studies found ring-averaged radial inflow velocities of $5-10$~\kms and $5-20$~\kms, respectively, though using different methods than in this study.    

In these models and observations, radial velocities of more than 40 \kms\ have not been observed; in our ring region in \gal\ these high values (see right $y$-axis in \autoref{fig:massflow_comp}) occur most likely due to differences in the adopted inclination values. The mass inflow rate is strongly dependent on the determined geometry of a galaxy, therefore, it is hard to make clear estimates of how many solar masses are transported to M83's central disk. However, we see ${\approx}1.9\times10^9\,M_{\odot}$ of gas at a distance of 16~kpc\footnote{Assuming that the ring mass (with average surface density ${\sim}3 \rm M_{\odot}/pc^2$ adopted from \autoref{fig:diff_reg}) is distributed over a 8~kpc wide area (i.e.\ the width of the ring, $r_{\rm gal}={\sim}8-16$~kpc).} from the center with a circular orbital period of $4.46\times10^3$~Myr\footnote{By taking the average $V_{\rm rot}$ at the ring radius ${\sim}160$ \kms\ from \autoref{fig:parameter_compare}.}. To reach inflow rates of of $10\,M_{\odot}$ per year would require infall on a timescale of ${\sim}2.0\times10^4$~Myr, which is 4 times an orbital time or 0.7 of a free-fall time\footnote{If we take $t_{\rm orb}{\sim}6~t_{\rm ff}$; see \autoref{app:time}}.

%
%
\section{Conclusion} 
\label{sec:Conclusions}
In this paper, we investigate the super-extended \hi\ disk of \gal\ based on 10 pointing VLA observations at 21$\arcsec$ ($\approx$~500pc) angular resolution. We find the following:
\begin{enumerate}
    \item[(1)] We detect significant \hi\ emission until a radius of ${\sim}$50~kpc. The most prominent features in \autoref{fig:mom0} are the northern extended arm, the southern arm, and the ring surrounding the central disk.
    
    \item[(2)] To investigate non-circular motions we adopted tilted model outputs from \cite{Heald2016} based on KAT7 observations. We find the highest deviations from pure circular motion in the ring (see \autoref{fig:mom1}).
    
    \item[(3)] We examine environmental differences in velocities (\vlos{} and \vres), atomic gas surface density (\sigatom), and velocity dispersion ($\sigma$). For that, we defined independent regions: central disk, ring, southern area, southern arm, and northern arm. In \autoref{fig:diff_reg} we find that overall \sigatom\ and \sig\ decrease with galactocentric radius, in agreement with prior work. The distribution of \vres\ towards the ring and northern arm shows a second peak $-25$ \kms.
    
    \item[(4)] We examine the radial and environmental dependence of velocity dispersion across the whole disk of \gal. We analyze the velocity dispersion, \sig\, using the effective width (\sigeff) and second moment (\sigmom) approach (see \autoref{fig:sig_rad}). We find enhanced velocity dispersion in the ring as well as the central disk using both techniques. This indicates that the atomic gas in the ring region seems to be more turbulent than in the outer regions.
    
    \item[(5)] We compare \sig\ with FUV emission to trace recent star formation, and do not find enhanced FUV emission where we find enhanced \hi\ velocity dispersion. This anti-correlation with star formation, suggest that radial transport could be the reason for enhanced velocity dispersion in the outskirts of \gal\ (see \autoref{fig:fuv_sigma}). 
    
    \item[(6)] We find an ``inflection point'' in observed velocities along the northern arm and speculate a possible connecting branch to the dwarf irregular galaxy UGCA\,365, that deviates from the general direction of the northern arm (see \autoref{fig:northernarm}).
    
    \item[(7)] We use a simplified model to get a radial mass flow rate profile and compare it with literature mass flow rates. We show that kinematic parameters have a strong impact on these profiles. Although mass inflow is one of the most important processes feeding star formation, tilted-ring models (2D or 3D) as the basis for deriving such profiles, are strongly sensitive on initial parameters (see e.g.\ \autoref{fig:massflow_comp}). Furthermore, warped or flaring \hi\ disks or distinct features at large radii complicate deriving kinematic parameters further. To investigate this in more detail, more sophisticated models are needed to correctly determine the kinematics of this extended structure.   
    
    \item[(8)] In agreement with recent studies, we find within the central disk, hints of inflowing material close to the nuclear region and outflowing material in the outer areas of the central disk. 
    
\end{enumerate}

%
\begin{acknowledgements}
  We would like to thank the anonymous referee for their insightful comments that helped improve the quality of the paper. 

  CE gratefully acknowledges funding from the Deutsche Forschungsgemeinschaft (DFG) Sachbeihilfe, grant number BI1546/3-1. 
  
  CE, FB, AB, IB, JdB and JP acknowledge funding from the European Research Council (ERC) under the European Union’s Horizon 2020 research and innovation programme (grant agreement No.726384/Empire). 
  
  The work of AKL is partially supported by the National Science Foundation under Grants No. 1615105, 1615109, and 1653300. 
  
  ER acknowledges the support of the Natural Sciences and Engineering Research Council of Canada (NSERC), funding reference number RGPIN-2017-03987.
  
  TGW acknowledges funding from the European Research Council (ERC) under the European Union’s Horizon 2020 research and innovation programme (grant agreement No. 694343)
  
  JMDK gratefully acknowledges funding from the European Research Council (ERC) under the European Union's Horizon 2020 research and innovation programme via the ERC Starting Grant MUSTANG (grant agreement number 714907). COOL Research DAO is a Decentralised Autonomous Organisation supporting research in astrophysics aimed at uncovering our cosmic origins.

  SCOG acknowledges funding from the European Research Council via the ERC Synergy Grant ``ECOGAL -- Understanding our Galactic ecosystem: From the disk of the Milky Way to the formation sites of stars and planets'' (project ID 855130), from the Deutsche Forschungsgemeinschaft (DFG) via the Collaborative Research Center (SFB 881 -- 138713538) ``The Milky Way System'' (subprojects A1, B1, B2 and B8) and from the Heidelberg Cluster of Excellence (EXC 2181 - 390900948) ``STRUCTURES: A unifying approach to emergent phenomena in the physical world, mathematics, and complex data'', funded by the German Excellence Strategy.
  
  MQ acknowledges support from the Spanish grant PID2019-106027GA-C44, funded by MCIN/AEI/10.13039/501100011033.
  
  K.G. is supported by the Australian Research Council through the Discovery Early Career Researcher Award (DECRA) Fellowship DE220100766 funded by the Australian Government.
  
  HH acknowledges the support of the Natural Sciences and Engineering Research Council of Canada (NSERC), funding reference number RGPIN-2017-03987 and the Canadian Space Agency funding reference 21EXPUVI3.
  
  EW is funded by the Deutsche Forschungsgemeinschaft (DFG, German Research Foundation) -- Project-ID 138713538 -- SFB 881 (``The Milky Way System'', subproject P2)
  
  WJGdB received funding from the European Research Council (ERC) under the European Union’s Horizon 2020 research and innovation programme (grant agreement No 882793 `MeerGas').

  D.J.P. is supported through the South African Research Chairs Initiative of the Department of Science and Technology and National Research Foundation and acknowledges partial support from NSF CAREER grant AST-1149491.  

  This work is based on observations carried out with the Karl G. Jansky Very Large Array (VLA) and the Green Bank Telescope (GBT). The National Radio Astronomy Observatory is a facility of the National Science Foundation operated under cooperative agreement by Associated Universities, Inc. The Green Bank Observatory is a facility of the National Science Foundation operated under cooperative agreement by Associated Universities, Inc. 
  
  This work is based in part on observations made with the \textit{Galaxy Evolution Explorer (GALEX)}. \textit{GALEX} is a NASA Small Explorer, whose mission was developed in cooperation with the Centre National d'Etudes Spatiales (CNES) of France and the Korean Ministry of Science and Technology. \textit{GALEX} is operated for NASA by the California Institute of Technology under NASA contract NAS5-98034.

  \textit{Software:} 
  \texttt{NumPy} \citep{NumPy2020},
  \texttt{SciPy} \citep{SciPy2020},
  \texttt{Astropy} \citep{Astropy2018},
  \texttt{pandas} \citep{Pandas_1.3.4},
  \texttt{Matplotlib} \citep{Matplotlib2007},
  \texttt{CASA} \citep{CASATeam2022},
  \texttt{BBarolo} \citep{EdT2021},
  \texttt{rotcur} \citep{Begeman1989},
  radio-astro-tools (\texttt{uvcombine}) \citep{Koch2022},
  \texttt{glnemo2} \url{https://projets.lam.fr/projects/glnemo2/wiki/}
  \texttt{astrometry} \url{https://nova.astrometry.net/}
\end{acknowledgements}

\bibliographystyle{aa}
\bibliography{references}

\appendix

\section{Image Combination}
\label{appendix:ImageCombination}
\begin{figure*}
    \centering
    \includegraphics[width=1.0\textwidth]{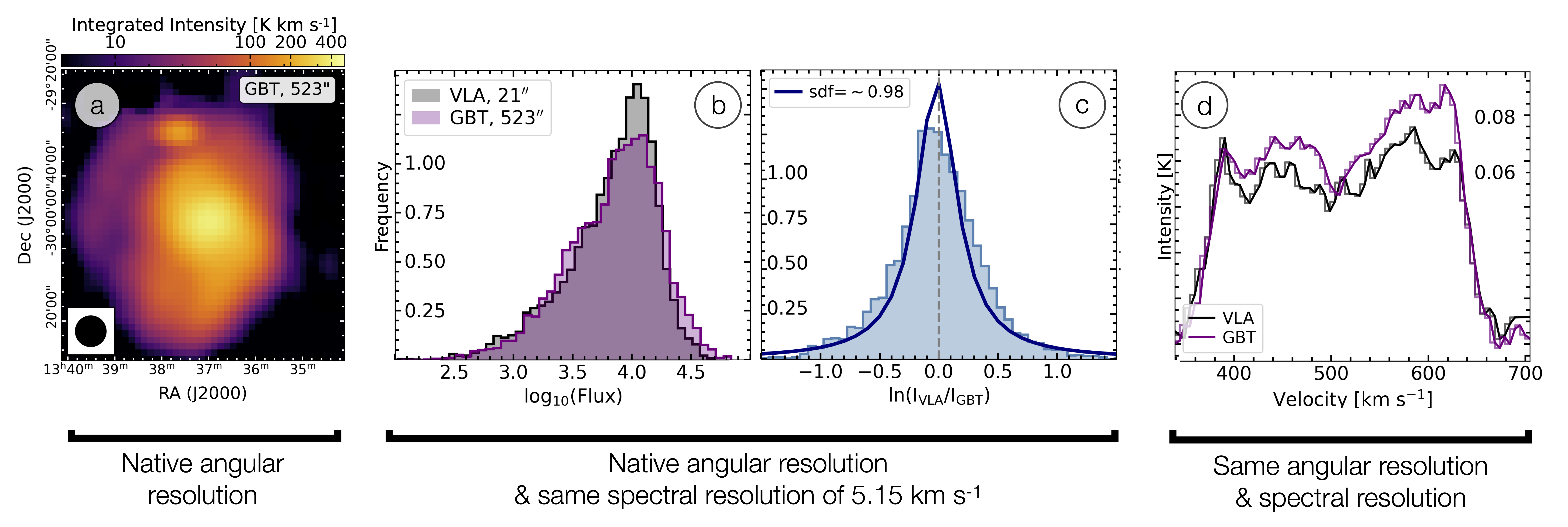}
    \caption{{Image combination.} \textbf{(a)} We show the integrated intensity maps of the GBT observation at native angular resolution. \textbf{(b)} Shows the distribution of the flux of the single dish (SD) and VLA data at matched spectral resolution (i.e. we regridded the VLA to the GBT spectral resolution).  \textbf{(c):} We used the \texttt{uvcombine} python package to find the single dish scaling factor (sdf; blue line). The ratio of the low-res and high-res observations result in a Cauchy distribution. We fit the log of the ratio to a Cauchy distribution (blue histogram) and its mean is then an estimate for the scaling factor. \textbf{(d)}: The spectra of the VLA and GBT observation at matched angular and spectral resolution (i.e. we gaussian convolve and regridded the VLA to the GBT angular and spectral resolution)}.
    \label{fig:feather_compare}
\end{figure*}

Similar to the procedure described in the Appendix of \cite{Koch2018}, itself based on chapter 3 of \cite{Stanimirovic1999}, we analyse the effects of combining the (i) VLA data with single dish (SD) (ii) Green Bank Telescope (GBT; project GBT11A-055) observations. 
(i) The VLA observation has an angular resolution of 21$\arcsec$ ($\sim500$\,pc) and a spectral resolution of 5.0 \kms\ . 
(ii) The GBT observation has an angular resolution of 523$\arcsec$ ($\sim12.13$\,kpc) and a spectral resolution of 5.15 \kms\ (see \autoref{fig:feather_compare}, panel 1a). 

We use a distribution-based method on the \uv-amplitudes where the spatial coverage of the VLA and SD data overlap to quantify the scaling factor. We use here the \texttt{uvcombine} package (\citealt{Koch2022}) and examine the distributions of the SD amplitudes, the VLA amplitudes, and their ratio across all channels. The amplitudes of the SD and VLA in the overlap regions are well-described by normal distributions, and as such, the ratio of the log-amplitude follows a Cauchy distribution \citep{Koch2018}. We fit a Cauchy distribution and use the location of the peak of the distribution as the scale factor. In \autoref{fig:feather_compare}, panel 1c, we show the distribution of the amplitudes across all channels and the best-fit Cauchy distribution (dark blue line). 

Adopting a maximum-likelihood approach and using the ratios from all channels we find a scaling factor of 0.982$\pm$0.003 for the GBT data. Because we expect the absolute flux uncertainties to be ${\sim}10\%$, the scaling factor from the distribution fit is consistent with 1 and thus we do not apply a scaling factor to the GBT data before feathering. 


\section{10 mosaic pointing and channel maps}
\label{appendix:add_figures}
\begin{figure}
    \centering
    \includegraphics[width=0.4\textwidth]{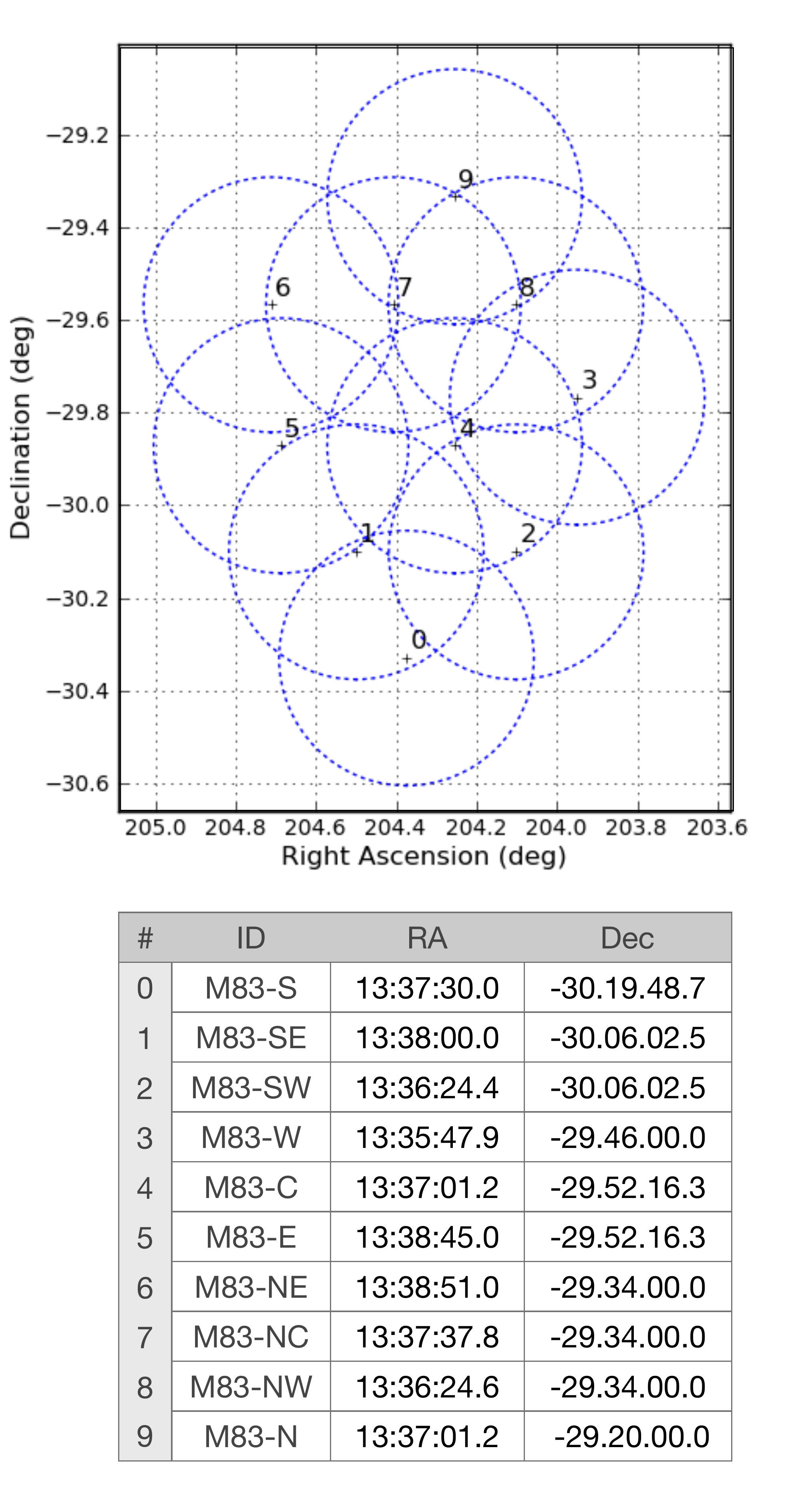}
    \caption{The position and coordinates for the extended ($\sim$1.5$^2$~deg) 10-point mosaic of our VLA observation.}
    \label{fig:pointing}
\end{figure}
\begin{figure*}[ht!]
    \centering
    \includegraphics[width=0.95\textwidth]{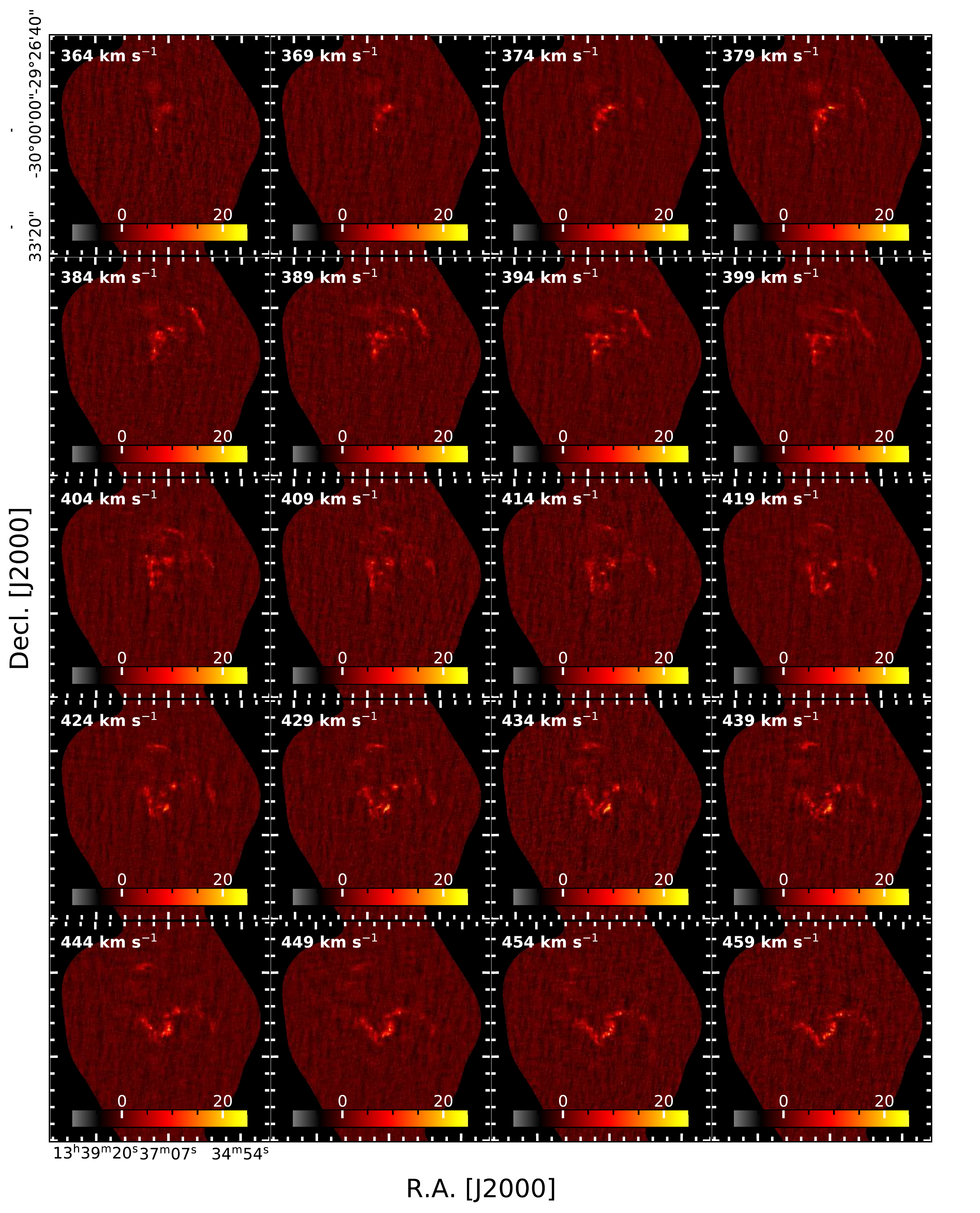}
    \caption{Channel maps of of our VLA+GBT data. Here we show now every 5.0 \kms\ channel. The line-of-sight velocity of the shown channels is displayed on the upper left corner of each panel.}
    \label{app:allchannels}
\end{figure*}

\begin{figure*}
    \centering
    \includegraphics[width=0.95\textwidth]{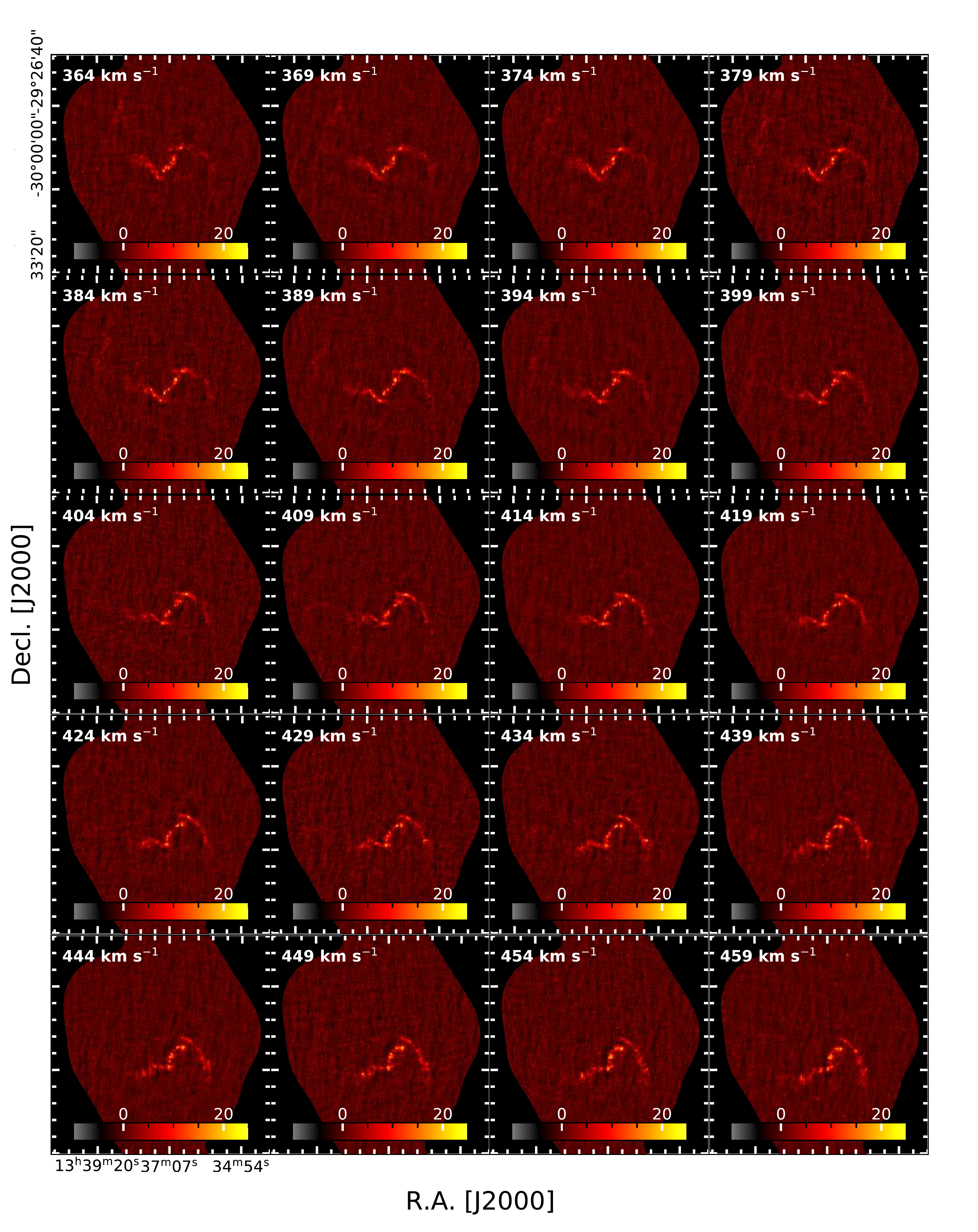}
    \caption{Fig. B.2. continued}
\end{figure*}

\begin{figure*}
    \centering
    \includegraphics[width=0.95\textwidth]{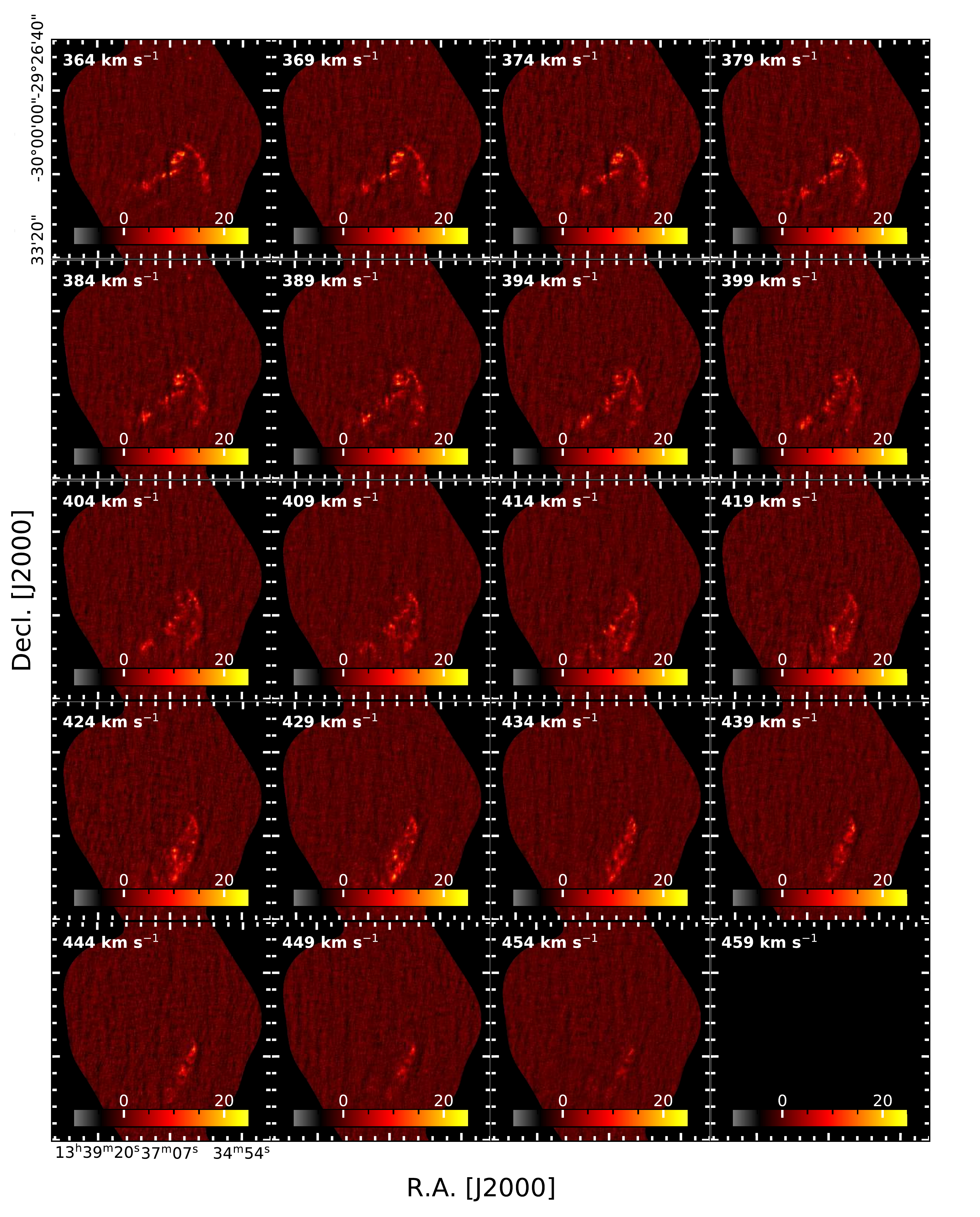}
    \caption{Fig. B.2. continued}
\end{figure*}

\autoref{fig:pointing} shows the 10 mosaic pointing for our VLA observation. We show the positions and their corresponding coordinates.

In \autoref{app:allchannels} we show all channels (contrary to the overview shown in \autoref{fig:chanmaps} where we took every third 5.0 \kms\ channel) of our imaged dataset. During the imaging procedure, we chose a spectral resolution of $\Delta_{v}=5.0 \,$\kms, to increase the signal-to-noise of our observations. \\

\section{ GBT observation}
\begin{figure}
    \centering
    \includegraphics[width=0.45\textwidth]{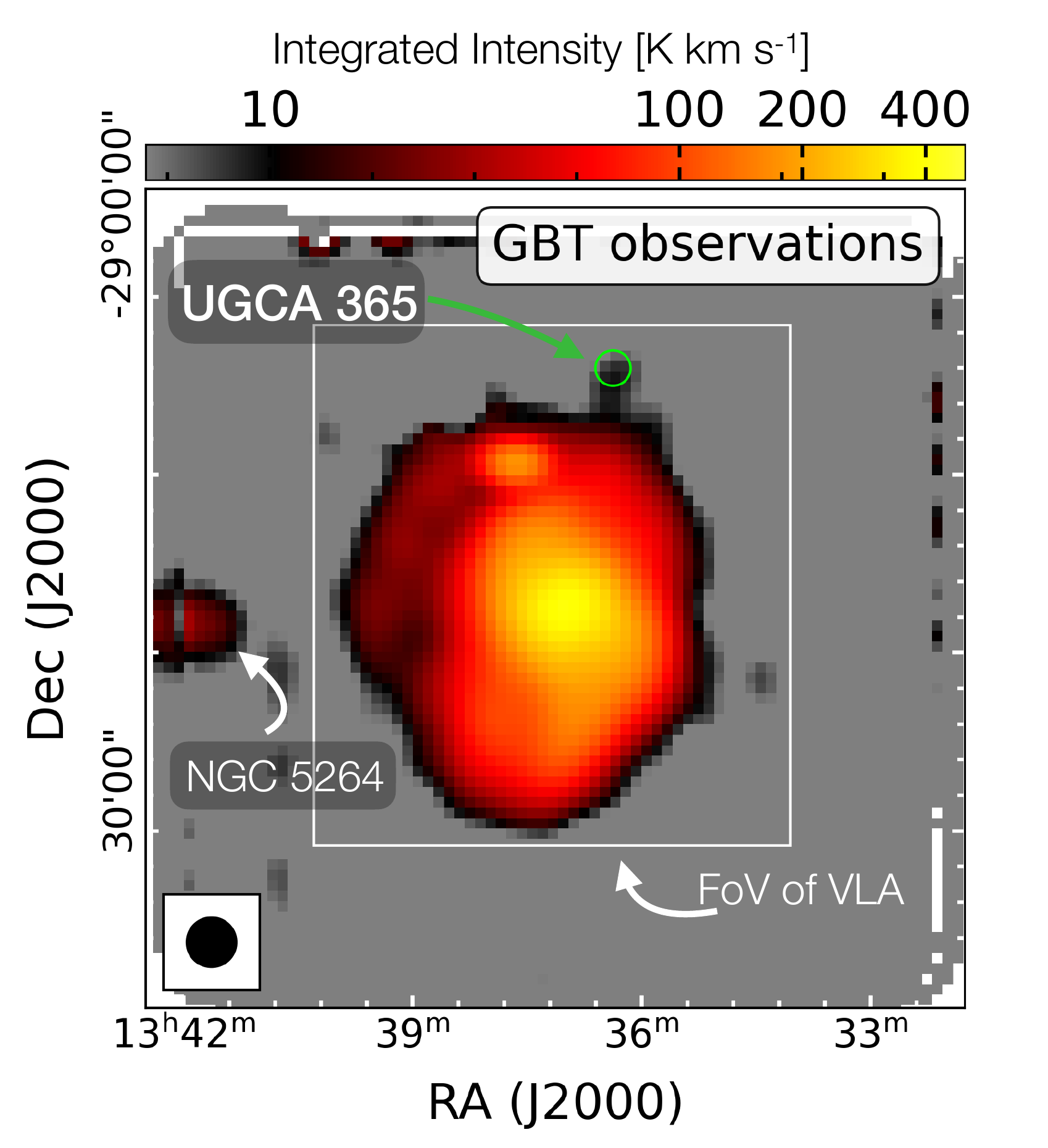}
    \caption{We show here the integrated intensity map of the GBT observation with white contours showing the field of view of the VLA observation. We denote the companion galaxies UGCA~365 and NGC~5264. The \hi\ emission of UGCA~365 is faint relative to \gal\ (${\sim}$10 K km s$^{-1}$ compared to ${\sim}$400 K km s$^{-1}$). }
    \label{fig:gbt}
\end{figure}

We show in \autoref{fig:gbt} the integrated intensity map of the GBT observation with angular resolution of 523$\arcsec$ together with the field of view of the VLA observation and the location of UGCA~365. The \hi\ emission of UGCA~365 is faint relative to \gal\ (${\sim}$10 K km s$^{-1}$ compared to ${\sim}$400 K km s$^{-1}$). 

\section{Multiple velocity components}
\begin{figure}
    \centering
    \includegraphics[width=0.5\textwidth]{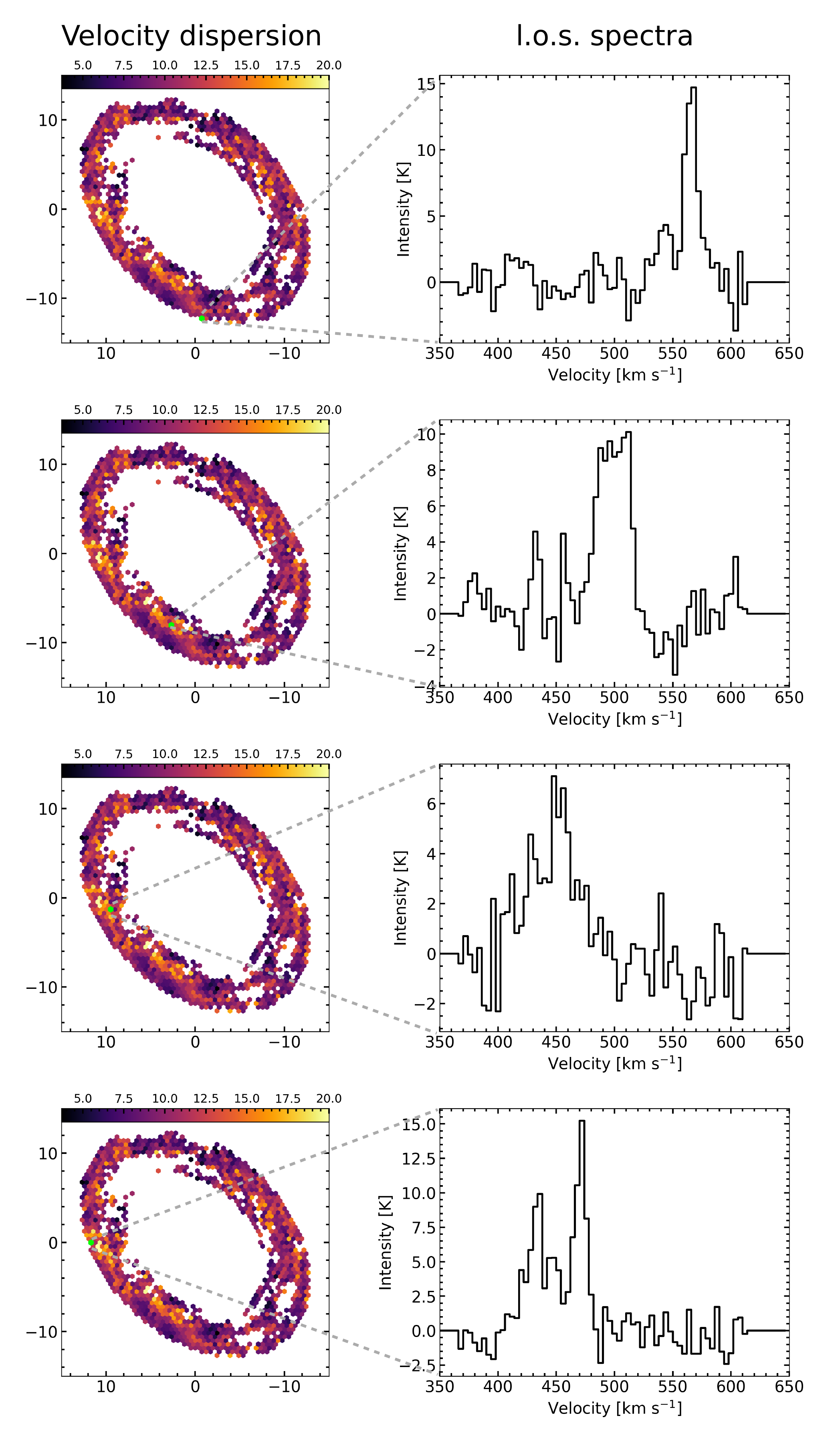}
    \caption{Examples of four spectra of individual line of sights that show higher values in the velocity map. These show the multi-component behavior of the spectra.}
    \label{app:spectra_ring_region}
\end{figure}
\begin{figure*}
    \centering
    \includegraphics[width=1.0\textwidth]{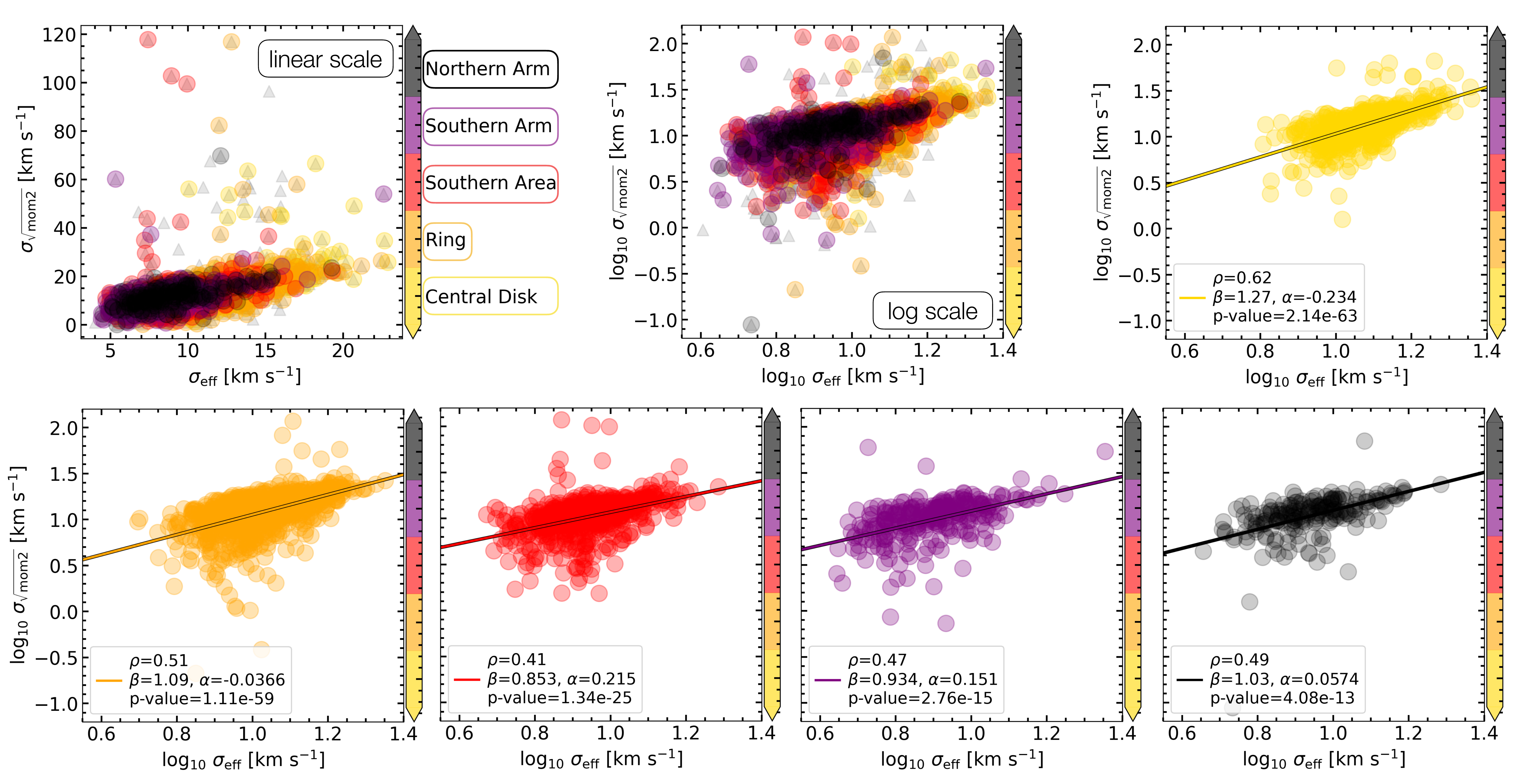}
    \caption{ Comparison of two approaches to determine the velocity dispersion. In all panels, the y-axes show the \sigmom\, and the x-axes the \sigeff\ values colored by the regions defined within this work (see the mask in \autoref{fig:diff_reg}). Only the upper left panel shows the x and y axes in linear scale, all the others are in log scale. We show linear regression fits between \sigmom\, and \sigeff\ for the central disk (yellow), ring (orange), southern area (red), southern arm (purple), and northern arm (black). We see the highest scatter in \sigmom\ in the ring region of ${\sim}2.5$dex. This is likely to represent a multi-component spectrum.}
    \label{fig:ew_vs_mom2}
\end{figure*}

In \autoref{sec:velocitydispersion} we mentioned that some of our sightlines show non-Gaussian or multiple velocity components. We show four examples in \autoref{app:spectra_ring_region}. We show in the left panels the velocity dispersion calculated with \autoref{eq:sigma} and show the hexagon in green, of which we show the spectrum on the right. These represent the multi-component behavior of the spectrum of individual lines of sight in the ring region.\\
We looked into individual regions and compared the two approaches discussed in this paper to derive the velocity dispersion -- \sigeff\ and \sigmom\ -- in \autoref{fig:ew_vs_mom2}. We show linear regression fits in the form of 
$\log(y) = \beta \times \log(x) + \alpha$. We find for the central disk and ring region a moderate\footnote{Here we refer to a moderate correlation if $\rho$ lies in the range of 0.5-0.7.} linear correlation between \sigeff\ and \sigmom. For all the other regions we find a weak, approximately linear correlation. We see the highest scatter in \sigmom\ in the ring region of ${\sim}2.5$~dex, followed by a scatter of ${\sim}2.0$~dex towards the central disk and southern area. These most likely highlight the multi-component and/or wider components of a spectrum. From the first panel, where we show the x and y axis in linear scales, we can see that velocity dispersions over ${\sim}20$ \kms\ are likely to represent a multi-component spectrum.

\section{ Orbital time and free fall time}\label{app:time}
In \autoref{sec:interpretingmassflowrates} we assumed that $t_{\rm orb}{\sim}6~t_{\rm ff}$. Here we show the derivation of that result. We assume that the material is distributed like a uniform density sphere. Then $t_{\rm ff}$ at some radius $R$ within that mass distribution is:
\begin{equation}
    t_{\rm ff} = \sqrt{\frac{3\pi}{32~G~\rho}},
\end{equation}
where $\rho$~=~M~/~((4~$\pi$~R$^{3}$)/3) where M is the mass enclosed within R and can also be expressed in terms of rotational velocities $v_{\rm rot}$:
\begin{equation}
    M = \frac{R~v_{\rm rot}^2}{G}.
\end{equation}
The orbital time $t_{\rm orb}$ is defined as:
\begin{equation}
    t_{\rm orb} = 2\pi\sqrt{\frac{R^3}{G~M}}.
\end{equation}
As a result we get:
\begin{equation}
    t_{\rm ff} = \frac{1}{2\sqrt{8}}~t_{\rm orb},
\end{equation}
or
\begin{equation}
    t_{\rm orb}{\sim}6~t_{\rm ff}
\end{equation}
This does, however, not imply that material \textit{will} undergo free-fall to the center since there is a lot of kinetic energy and angular momentum that will prevent this. It nevertheless serves as a good estimate.

\end{document}